\documentclass[aps,pra,twocolumn,superscriptaddress]{revtex4-2}
\usepackage{graphicx}
\usepackage{url}
\usepackage{amsmath,amsthm,amssymb,physics,color,enumerate,tabularx,makecell,mathrsfs,float,bbm,stackrel,mathtools,multirow,bm}
\usepackage{natbib}
\usepackage{ulem}

\usepackage[colorlinks=true,%
bookmarks=false,%
linkcolor=blue,%
urlcolor=blue,%
citecolor=blue,%
breaklinks]{hyperref}

\usepackage{soul} 

\usepackage[dvipsnames]{xcolor}

\newcommand{\eref}[1]{(\ref{#1})}

\begin{document}
\title{
Learning feedback control strategies for quantum metrology
}
\author{Alessio Fallani}
\affiliation{Dipartimento di 
Fisica ``Aldo Pontremoli'', Universit\`{a} degli Studi di Milano, I-20133 Milano, Italy}

\author{Matteo A. C. Rossi}
\affiliation{InstituteQ - the Finnish Quantum Institute, Aalto University, Finland}
\affiliation{QTF Centre of Excellence, Department of Applied Physics, Aalto University, FI-00076 Aalto, Finland}
\affiliation{Algorithmiq Ltd., Kanavakatu 3C, FI-00160 Helsinki, Finland}

\author{Dario Tamascelli}
\affiliation{Dipartimento di 
Fisica ``Aldo Pontremoli'', Universit\`{a} degli Studi di Milano, I-20133 Milano, Italy}

\author{Marco G. Genoni}
\affiliation{Dipartimento di 
Fisica ``Aldo Pontremoli'', Universit\`{a} degli Studi di Milano, I-20133 Milano, Italy}
\begin{abstract}
We consider the problem of frequency estimation for a single bosonic field evolving under a squeezing Hamiltonian and continuously monitored via homodyne detection. In particular, we exploit reinforcement learning techniques to devise feedback control strategies achieving increased estimation precision. We show that the feedback control determined by the neural network greatly surpasses in the long-time limit the performances of both the ``no-control'' strategy and the standard ``open-loop control'' strategy, which we considered as benchmarks. We indeed observe how the devised strategy is able to optimize the nontrivial estimation problem by preparing a large fraction of trajectories corresponding to more sensitive quantum conditional states.
\end{abstract}

\date{\today}
\maketitle

\section{Introduction}
\normalem
The goal of quantum metrology is to devise strategies able to exploit purely quantum properties, such as entanglement and squeezing, in order to estimate parameters with a precision beyond the one obtainable via classical means \cite{GiovannettiNatPhot,PirandolaReview}. In the classical domain it is usual to study estimation strategies based on the continuous monitoring of a system, leading to sensors that have applications ranging from engineering to medicine. 

This kind of approach is particular interesting in the context of quantum metrology with continuously monitored quantum systems~\cite{WisemanMilburn,SteckJacobs}. The role of continuous measurements is indeed twofold: on the one hand, as it happens { classically}, the measurement output is exploited to acquire information on the parameters characterizing the system; on the other, the act of measuring alters the state of the system itself, thus opening the possibility of dynamically prepare more sensitive quantum probes. Several works have been proposed in the literature, both discussing the fundamental statistical tools to assess the precision achievable in this framework~\cite{Guta2007a,Tsang2011,Tsang2013a,GammelmarkCRB,GammelmarkQCRB,Guta2016,Genoni2017,Albarelli2017a}, and presenting practical estimation strategies~\cite{Mabuchi1996,Gambetta2001,Geremia2003,Molmer2004,Madsen2004,Stockton2004,Tsang2010,Wheatley2010,Yonezawa2012,Cook2014,Six2015,KiilerichHomodyne,Cortez2017,Ralph2017,Atalaya2017,Albarelli2018Quantum,Shankar2019,Rossi2020PRL}. 

Moreover, in the context of continuously monitored quantum systems, it is also natural to study strategies able to exploit this information in order to steer the evolution towards a desired quantum state via feedback control~\cite{WisemanMilburn,Doherty1999}. Much effort has been devoted to the design of strategies able to generate metrologically relevant quantum states, such as squeezed states ~\cite{Wiseman1993,Wiseman1994,Thomsen2002,SerafozziMancini,Szorkovszky2011,Genoni2013PRA,Genoni2015NJP,Hofer2015,Brunelli2019PRL,DiGiovanni2021}, or to cool optomechanical systems towards their ground state, with the outstanding experimental results recently observed in Refs.~\cite{Rossi2018,Magrini2021,Tebbenjohanns2021}.
\par
Reinforcement learning (RL) is one of the main paradigms of machine learning, together with supervised and unsupervised learning. In RL an agent learns how to perform a task by acting on a system and updating its policy through a reward/punishment mechanism \cite{sutton2018reinforcement}. The introduction of deep neural networks in RL has led to formidable results: machines have, e.g., learned how to play video games \cite{deepmindgames2015} or how to beat expert human players at complex board games like Go \cite{go2017nature}. 

RL has been recently applied in the context of quantum information, and more in general to quantum technology, to find optimal strategies for some designated tasks~\cite{MarquardtTutorial}, ranging from optimizing feedback for quantum error correction \cite{Fosel2018} to quantum control strategies \cite{Mavadia2017,Niu2019}, and from optimizing quantum transport~\cite{Porotti2019,Brown2021} to quantum compiling \cite{Moro2021}, or even to solve the Rubik's cube by exploiting quantum mechanics~\cite{Corli2021}. Recently, RL has been also employed to optimize feedback control protocols in continuously monitored quantum systems \cite{borah2021,Porotti2021,evans2021stochastic}, with a main focus on quantum state engineering. 

Discovering feedback strategies, where decisions are based on previously observed measurement results, is indeed a challenging task. The stochastic nature of the problem, together with the presence of feedback mechanisms, leads to a doubly-exponential growth of the space of possible strategies with respect to the number of time steps. Such a task falls therefore beyond the scope of standard optimal control, and also supervised learning, techniques \cite{Porotti2021}. On the other end, it suits the RL paradigm: the agent explores the problem space by performing random experiments on the system while learning, at the same time, an action policy.

In this work we exploit RL to design a feedback strategy optimizing a given non trivial metrological problem. In particular, we consider the estimation of the frequency of a harmonic oscillator subjected to a squeezing Hamiltonian and undergoing a continuous homodyne detection. Differently form previous work \cite{borah2021,Porotti2021,evans2021stochastic}, where the goal was the preparation of a given target state to be exploited in a selected quantum information protocol, in this paper we aim to optimize real-time feedback for quantum metrology purposes,  as to  attain a high precision in parameter estimation without targeting the preparation of a precise quantum state. 

We show that the feedback strategy determined by RL provides a high precision in parameter estimation, and overcomes the performance of some benchmark approaches. Interestingly enough, the feedback protocol determined by the agent optimizes the interplay between the squeezing direction and the displacement. Given the stochastic nature of the dynamics induced by the measurement back-action, such a strategy is highly non-trivial and can not be easily obtained with standard optimal control techniques or supervised learning techniques.

\par
The manuscript is organized as follows: In Sec. \ref{s:model} we present the physical model and the estimation problem; in Sec. \ref{s:estimation} we introduce the figures of merit that we will employ to assess the protocols and we discuss the role of squeezing and feedback in the estimation procedure. In Sec. \ref{s:RL} we show how we apply RL to our problem and in Sec. \ref{s:results} we present our main results. We conclude our manuscript in Sec. \ref{s:conclusion} with a brief discussion and some outlooks.
%
%
\section{The estimation problem}
\label{s:model}
We consider a single bosonic mode described by the quadrature operators $(\hat{q},\hat{p})$ satisfying the canonical commutation relation $[\hat{q},\hat{p}]=i\mathbbm{1}$~\cite{SerafozziBook}. The evolution of the mode is determined by the Hamiltonian
\begin{align} 
\hat{H}_0 = \omega \, \hat{a}^\dag \hat{a} + \chi (\hat{a}^2 + \hat{a}^{\dag 2} ) \,
\label{eq:Hamiltonian}
\end{align}
with $\hat{a} = (\hat{q} + i \hat{p})/\sqrt{2}$ denoting the annihilation operator. The first term simply corresponds to the usual free quantum oscillator Hamiltonian, characterized by a frequency $\omega$; the second is a single-mode squeezing term able to generate, for $\omega=0$ and $\chi>0$, squeezing in the $\hat{q}$ quadrature. We remind here that a quantum state is said to be squeezed if it presents fluctuations of a quadrature operator below the vacuum shot-noise. The amount of squeezing of a quantum state $\varrho$, for example for the $\hat{q}$ quadrature, is typically evaluated in dB according to the formula 
\begin{align}
    \xi_{\sf dB} =-10 \log_{10} (\langle \Delta \hat{q}^2 \rangle/\langle \Delta \hat{q}^2 \rangle_{0}) \,,
\end{align} where we have denoted with $\langle \Delta \hat{q}^2 \rangle$ and $\langle \Delta \hat{q}^2 \rangle_{0} =1/2$ the variance of $\hat{q}$ evaluated respectively for the quantum state $\varrho$ and for the vacuum state $|0\rangle$. More in general, the maximum amount of squeezing of a single-mode quantum state along a generic quadrature operator can be evaluated as $\xi_{\sf dB} =-10 \log_{10} (\lambda_-)$, where $\lambda_-$ denotes the minimum eigenvalue of its covariance matrix $\boldsymbol{\sigma}$ (see more details on covariance matrices and Gaussian formalism in Appendix \ref{a:GaussianMonitored}).

Physically, the Hamiltonian (\ref{eq:Hamiltonian}) describes an optical parametric oscillator (OPO) that is a cavity mode with resonance frequency $\omega_c$ interacting with a nonlinear crystal and driven by a laser with frequency $\omega_l$. It can be indeed obtained by going to a frame rotating at the laser frequency, with $\omega = \omega_c - \omega_l$ denoting the detuning between cavity resonance and laser. In what follows we focus on the problem of estimating the fixed, but unknown, value of this detuning parameter $\omega$. This kind of estimation problem has been recently discussed in the standard open-system scenario for a circuit-QED implementation, by also considering the usefulness of an extra Kerr-type nonlinearity in \cite{Dicandia2021}. We remark that in the continuous-variable scenario one typically considers the estimation of an optical phase accumulated during a finite time evolution~\cite{Monras2006,GenoniPRL2011}. Phase estimation and frequency estimation are, however, fundamentally equivalent and we will focus on the latter as in our setup we have to deal with with a time-continuous evolution. While we phrase our results in terms of a quantum optical scenario, we expect that our findings can be extended to other physical platforms where frequency estimation is at the basis of quantum enhanced atomic clocks~\cite{SchleierSmith2010,Leroux2010} and quantum magnetometry~\cite{Wasilewski2010}.

In our setting, the cavity mode is subjected to loss at rate $\kappa$. The output (leaking field) signal is then measured by means of a continuous homodyne measurement, performed with efficiency $\eta$ (this parameter $\eta=\eta_D \eta_L$ takes into account both the homodyne detector efficiency $\eta_D$ and the fraction of output field $\eta_L$ that is not collected by the detector). The corresponding continuous measurement outcome can be written as
\begin{align}
dy_t = \sqrt{\eta \kappa} \langle \hat{a} + \hat{a}^\dag \rangle_c \,dt + dw_t, \label{eq:photocurrent1}
\end{align}
where $\langle \cdot \rangle_c = \Tr[\varrho_c \cdot]$ denotes the expectation over the conditional state $\varrho_c$, and $dw_t$ is a Wiener increment, characterized by $\mathbbm{E}[dw_t]=0$  and $\mathbbm{E}[dw_t^2]=dt$. 

Under these assumptions the evolution of the conditional state is governed by the stochastic master equation \cite{WisemanMilburn}
\begin{align}
d\varrho_c &= - i [\hat{H}_0, \varrho_c] \,dt + \kappa \mathcal{D}[\hat{a}]\varrho_c \,dt \nonumber \\
& + \sqrt{\eta \kappa}\mathcal{H}[\hat{a}] \varrho_c \,dw_t \,, \label{eq:sme}
\end{align}
where
\begin{align}
\mathcal{D}[\hat{a}]\varrho_c &= \hat{a} \varrho_c \hat{a}^\dag - \frac{\hat{a}^\dag \hat{a} \varrho_c + \varrho_c \hat{a}^\dag \hat{a}}{2}\,, \\
\mathcal{H}[\hat{a}]\varrho_c &= \hat{a} \varrho_c + \varrho_c \hat{a}^\dag - \langle \hat{a} + \hat{a}^\dag \rangle_c\, \varrho_c \,.
\end{align}
Notice that the sequence of measures $\tilde{y}_t = \{ dy_s \}_{s=0}^{t} ,\ 0 \leq s \leq t$ determines the trajectory followed by the conditional state $\varrho_c$ up to time $t,$ and that the value of $\omega$ determines the conditional joint probability density $p(\tilde{y}_t|\omega)$.

In particular, the stochastic master equation (\ref{eq:sme}) for the conditional state $\varrho_c$ is completely equivalent to the equations for its first moments vector $\bar{\bf r}_c$ and covariance matrix $\boldsymbol{\sigma}_c$ \cite{WisemanDoherty,Diffusione,SerafozziBook}
\begin{align}
d\bar{\bf r}_c &= A \bar{\bf r}_c \,dt + (E - \boldsymbol{\sigma}_c B) \frac{{\bf dw}_t}{\sqrt{2}} \,, \label{eq:rc} \\
\frac{d\boldsymbol{\sigma}_c}{dt} &= A \boldsymbol{\sigma}_c + \boldsymbol{\sigma}_c A^{\sf T} + D - (E - \boldsymbol{\sigma}_c B)(E - \boldsymbol{\sigma}_c B)^{\sf T}.
\label{eq:sigmac}
\end{align}

The continuous measurement outcome (\ref{eq:photocurrent1}) can be written in vectorial form as ${\bf dy}_t = -\sqrt{2} B^{\sf T} \bar{\bf r}_c \,dt + {\bf dw}_t$\, with ${\bf dw}_t$ the vector of uncorrelated Wiener increments entering also in Eq. (\ref{eq:rc}). We refer the reader to Appendix \ref{a:GaussianMonitored} for details on the matrices entering in Eqs. \eref{eq:rc} and \eref{eq:sigmac}. 

It is important to remark  here that the dynamics determined by the above equations is stable, i.e. leads to a steady-state, if and only if the Hurwitz condition ${\rm Re}[{\rm eigs}(A)]<0$ is satisfied, that is if the the real part of the eigenvalues of the drift matrix $A$ is strictly smaller than zero. In our case it corresponds to the inequality $\chi < |\kappa/2|$ and we will always assume that this condition is fulfilled.

As we pointed out before, the Hamiltonian in Eq. (\ref{eq:Hamiltonian}) is able to generate squeezing. If we focus on the unmonitored (unconditional) dynamics (i.e. for $\eta=0$), the maximum squeezing at steady-state is obtained in the case of $\omega=0$, leading to a steady-state variance of the $\hat{q}$ quadrature $\langle \Delta \hat{q}^2 \rangle_{\sf unc} = \kappa/(2\kappa + 4 \chi)$, that is indeed below the vacuum limit $\langle \Delta \hat{q}^2 \rangle_{0} =1/2$ for $0<\chi<\kappa/2$ (for negative values of $\chi$ one would obtain squeezing along the $\hat{p}$ quadrature). We observe, in particular, that the squeezing increases approaching instability and that for $\chi \approx \kappa/2$ the well known limit of $\xi_{\sf dB} = 3$dB of squeezing is saturated~\cite{MilburnWallsSqueezing,CollettGardinerSqueezing}. 
The Riccati equation (\ref{eq:sigmac}) can be analytically solved in this case and its solution shows that continuous monitoring allows indeed to greatly enhance the squeezing generation for the conditional states $\varrho_c$. In particular, at steady-state and for $\eta=1$, a variance  $\langle \Delta \hat{q}^2 \rangle_{\rm c} = (\kappa -2 \chi)/(2\kappa)$ is obtained, thus approaching infinite squeezing near criticality. 

For $\omega \neq 0$, no analytical solution is available, but we find by numerical means that a smaller, but still beyond the $3$dB limit, amount of squeezing can be obtained at steady-state; in this case, moreover, the maximum value  of the squeezing corresponds, in general, to quadratures different from $\hat{q}$ and $\hat{p}$. 

\section{Frequency estimation, squeezing and feedback optimization}
\label{s:estimation}
Our goal is to devise a protocol able to estimate the frequency parameter $\omega$ with high precision. In particular we will compare  the performance achieved by our proposal with those of different alternative strategies that will be detailed later in the manuscript. In all these strategies, information on the unknown parameter is going to be obtained form two sources: the continuous measurement outcome $y_t$ and a final strong measurement on the corresponding conditional states $\varrho_c$. 

The observation above is made rigorous by observing the form of the corresponding quantum Cram\'er-Rao bound that applies in this scenario. As customary in the context of frequency estimation, we will consider the total time of the experiment $T$, divided in $M$ single runs of duration $t=T/M$ as a fixed resource \cite{Huelga97}. Under this assumption, one proves that the precision $\delta \omega$ of any possible unbiased estimator is lower bounded as \cite{Albarelli2018Quantum}
\begin{align}
    \delta \omega \, \sqrt{T} \geq \frac{1}{\sqrt{\mathcal{Q}_{\sf eff}/t}} \,, \label{eq:QCRB}
\end{align}
where we have defined the {\em effective} quantum Fisher information \cite{Albarelli2017a,Albarelli2018Quantum}
\begin{align} 
\mathcal{Q}_{\sf eff} = \mathcal{F}_{\sf hom}  + \bar{\mathcal{Q}}_c
\,. \label{eq:effQFI}
\end{align}
The first term, defined as $\mathcal{F}_{\sf hom}=\mathcal{F}[p(\tilde{y}_t|\omega)]$, does indeed correspond to the classical Fisher information of the conditional probability of observing a trajectory given the value of the parameter $\omega$, and thus to the information obtainable via the continuous homodyne detection \cite{Gammelmark2013a,Genoni2017}. The second term, that we define as $\bar{\mathcal{Q}}_c = \mathbbm{E}_{\sf traj}\left[\mathcal{Q}[\varrho_c] \right]$, is the average of the quantum Fisher information (QFI) $\mathcal{Q}[\varrho_c]$ of the different conditional states generated by the measurement: it thus quantifies the average information obtainable via a final measurement on the different trajectory-dependent $\varrho_c$ (we have introduced the notation $\mathbbm{E}_{\sf traj}[\cdot]$ to denote the average over the conditional distribution $p(\tilde{y}_t | \omega)$ of the different trajectories defined by the stream of measurement outcomes $\tilde{y}_t$). Both these quantities can be numerically obtained via the evolution of the first and second moments of the conditional states ${\bf r}_c$, $\boldsymbol{\sigma}_c$, via their derivatives with respect to the parameter, i.e. $\partial_\omega {\bf r}_c$ and $\partial_\omega\boldsymbol{\sigma}_c$, and by performing a Monte Carlo average of the trajectories (see Appendix \ref{a:QMetrCM} for more details). According to Eq. (\ref{eq:QCRB}), the quantity $\mathcal{Q}_{\sf eff}/t$ will thus act as our figure of merit to assess the different estimation protocols that we will discuss in the next sections. 
\par
The feedback strategy we are going to consider later on exploits the information obtained from the continuous measurement output $dy_t$ to perform a unitary feedback operation \cite{WisemanMilburn} via the Hamiltonian $\hat{H}_{\sf fb} = \omega_{\sf fb}(t) \,\hat{a}^\dag \hat{a}$, that is by either changing the laser or the cavity resonance frequency via the (possibly time-dependent) parameter $\omega_{\sf fb}(t)$.  

In order to better understand the motivation of a machine learning approach for the optimization of such feedback strategy, it is expedient to discuss the peculiar features of the estimation problem we are considering. As we have discussed in Sec. \ref{s:model}, via Hamiltonian $\hat{H}_0$ in Eq. (\ref{eq:Hamiltonian}), by fixing $\omega=0$ and by assuming a positive coupling $\chi>0$, we know that unconditional squeezing is generated for the quadrature $\hat{q}$, and that a maximum of $3$dB can be obtained at steady-state near instability (that is for $\chi \approx \kappa/2$)~\cite{MilburnWallsSqueezing,CollettGardinerSqueezing}. However, if we also include a continuous homodyne detection, as the one described by the stochastic master equation (\ref{eq:sme}), the squeezing of the conditional states can be greatly enhanced, going well beyond the $3$dB limit. 

\begin{figure}[t]
    \centering
    \includegraphics[width=\columnwidth]{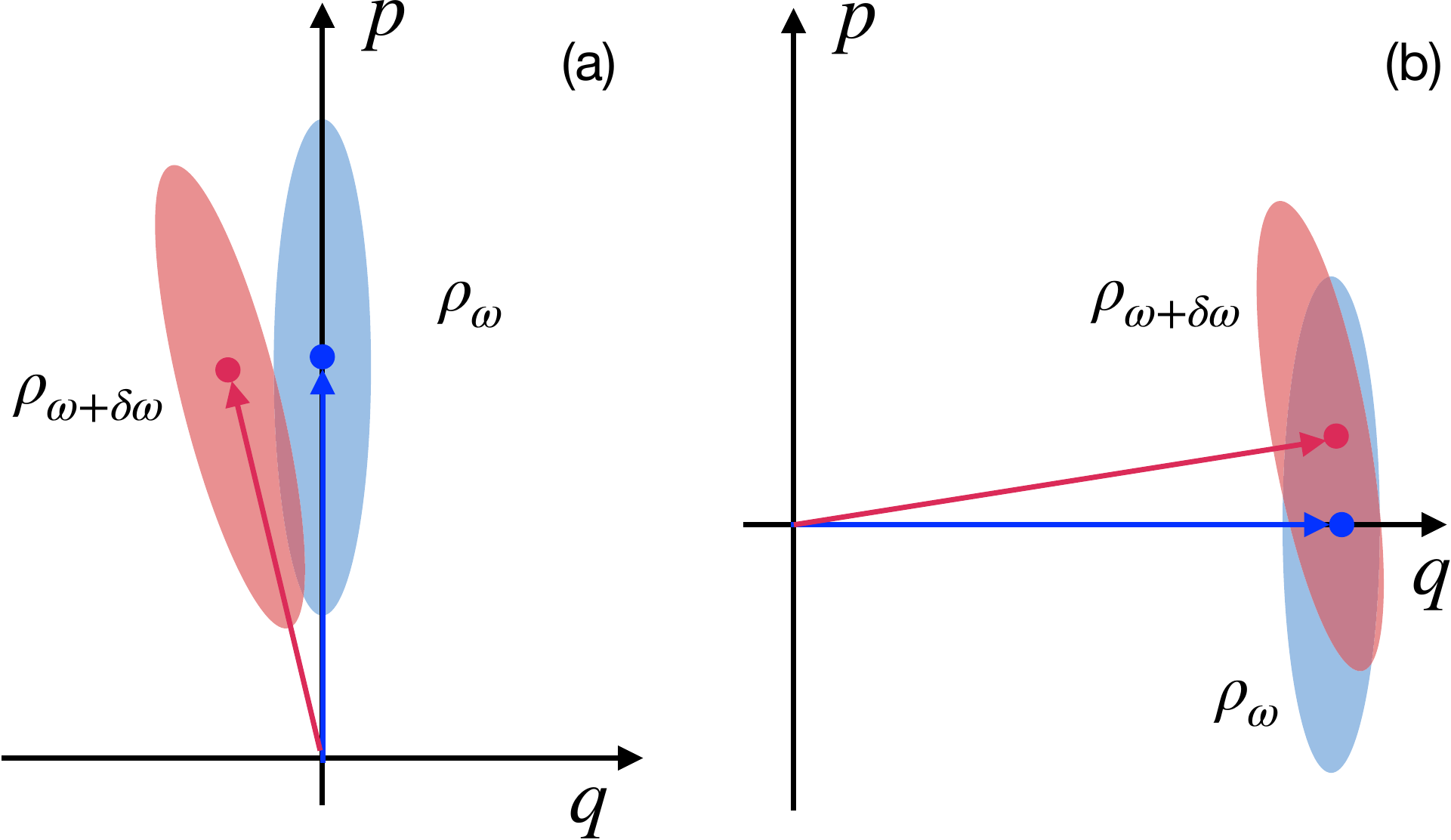}
    \caption{We show a heuristic representation in phase-space of frequency (phase) estimation via displaced squeezed state. The estimation precision on the parameter $\omega$ can indeed be understood both qualitatively and quantitatively \cite{MatteoIJQI} in terms of the distinguishability of quantum states $\varrho_\omega$ and $\varrho_{\omega+\delta\omega}$ differing by an infinitesimal value $\delta\omega$ of the parameter. In panel (a) we observe that, if the state is displaced along the $\hat{p}$ axis of the phase space and it is squeezed along $\hat{q}$, the two states are highly distinguishable, and thus $\omega$ can be measured with high precision. However, as one can see in panel (b), the quantum states become very indistinguishable in the opposite case, that is for both displacement and squeezing along $\hat{q}$. Clearly the situation is much worse for large values of $|\bar{\bf r}_c|$, while for small first moments, the benefit of squeezing may still yield a high estimation precision.}
    \label{f:phaseest}
\end{figure}

The continuous monitoring has however also another effect on the conditional state, that is, it gives a stochastic nonzero value for the first moments as described in Eq. (\ref{eq:rc}). Squeezing and nonzero first moments are the relevant figures of merit for the estimation problem we are indeed considering. Squeezing by itself is typically the most important resource for frequency estimation (or analogously for phase estimation \cite{Monras2006,GenoniPRL2011}). However its interplay with nonzero first moments may play a crucial role in determining the estimation precision. A heuristic representation of this fact is given in Fig. \ref{f:phaseest}: we observe that squeezing could further enhance the estimation if the squeezed quadrature is orthogonal to the direction of the first moment vector $\bar{\bf r}_c$ in phase space. Remarkably, if $|\bar{\bf r}_c |$ is large enough, squeezing for the quadrature parallel to the direction of $\bar{\bf r}_c$ is going to be detrimental for the estimation of $\omega$. We remark that in fact there is a non-trivial trade-off between the amount of squeezing and $|\bar{\bf r}_c|$, as for small enough $|\bar{\bf r}_c|$, squeezing along the {\em wrong} direction is still going to be a useful resource for estimation. We thus expect that the RL agent will be able to optimize such nontrivial problem, by devising feedback strategies able not only to generate large squeezing but also to generate non-zero first moments, and, more importantly to adjust their relative directions in phase space. We remark that in general a final non-Gaussian measurement on the conditional states may be needed in order to saturate the corresponding quantum Cram\'er-Rao bound. However, as demonstrated in phase-estimation protocols with Gaussian states \cite{Monras2006,Oh2019} and as we will describe in Appendix \ref{a:finalhomodyne} for our results, a final homodyne detection is going to extract in general a fair amount of the maximum amount of information, being nearly optimal for pure Gaussian states.
\section{Applying Reinforcement Learning}
\label{s:RL} 
As we mentioned in the Introduction, RL deals with a reward-based learning paradigm. An agent learns how to achieve a certain goal by performing actions on an {environment}, obtaining complete or partial information on its {state}, and a reward, specifically designed for the goal.

In this framework the agent is trained over a number of simulations with finite duration called episodes. At each timestep $dt$, we give the agent access to all the possible information of the conditional state and on its dependence on the parameter $\omega$, that is by considering as observations the set of parameters ${\bf obs} = \left( \bar{\bf r}_c, \boldsymbol{\sigma}_c, \partial_\omega \bar{\bf r}_c, \partial_\omega \boldsymbol{\sigma}_c, {\bf dy}_t \right)$. We remark that all these quantities can be updated at each time, according to Eqs. (\ref{eq:rc}), (\ref{eq:sigmac}), (\ref{eq:drc}), (\ref{eq:dsigmac}) once the continuous measurement result ${\bf dy}_t$ is obtained. The agent then performs
an action on the environment, which in our case consists directly in the choice of a real value for the feedback parameter $\omega_{\sf fb}$.

One of the most important steps in defining a RL problem is to identify the correct reward function. As discussed in the previous section, we will assess our feedback strategies via the effective QFI per time $\mathcal{Q}_{\sf eff}/t$. We first observe that the Fisher information corresponding to the continuous homodyne detection can be written as (more details in Appendix \ref{a:GaussianMonitored})
\begin{align}
    \mathcal{F}_{\sf hom} = \mathbbm{E}_{\sf traj}\left[2 \int dt \, (\partial_\omega \bar{\bf r}_c)^{\sf T} B B^{\sf T} (\partial_\omega \bar{\bf r}_c)\right] \,. \label{eq:Fhom}
\end{align}
As a consequence we may write
\begin{align}
    \frac{\mathcal{Q}_{\sf eff}}{t} = \mathbbm{E}_{\sf traj}\left[ \mathcal{R} \right]
\end{align}
where we have defined a (positive) trajectory dependent quantity
\begin{align}
    \mathcal{R} = \frac{2\int dt \, (\partial_\omega \bar{\bf r}_c)^{\sf T} B B^{\sf T} (\partial_\omega \bar{\bf r}_c) + \mathcal{Q}[\varrho_c]}{t} \,.
\end{align}

This observation allows us to state that the maximization of $\mathcal{Q}_{\sf eff}/t$ corresponds to the {\em trajectory-wise} maximization of $\mathcal{R}$ that will thus act as our reward function (we remind the fact that, as described in Appendix \ref{a:GaussianMonitored}, $\mathcal{Q}[\varrho_c]$ can be easily evaluated from the properties of the Gaussian conditional state $\varrho_c$).

We here used the algorithm Proximal Policy Optimization (PPO) \cite{schulman2017proximal}, a state-of-the-art actor-critic algorithm where the agent is a neural network optimizing both its evaluation of the future reward (critic) and its reward maximization strategy (actor). We exploited the implementation of PPO available in the package \textsc{stable-baselines} \cite{stable-baselines}. For this algorithm the strategy, also called policy, is a stochastic one, meaning that the action of the agent is extracted from a Gaussian distribution. 

The agent we trained is a neural network with a feed-forward and fully connected architecture, composed by an input layer of the size of the observations connected to two distinct $64 \times 64$ networks (one for the actor and one for the critic). The network is trained using a gradient descent method with linearly decreasing learning rate starting from a value of $2.5\cdot 10^{-4}$ and an entropy coefficient of 0.001 and a discount factor $\gamma = 0.99$. At every step of training the loss function is evaluated on on batches of 512 elements given by the experience of four parallel workers over a time horizon of 128 timesteps. The total number of timesteps included in the training is $30 \cdot 10^6$ composed by consecutive simulations (episodes) with finite duration of $10^5$ steps. At the beginning of each episode the initial condition for the system is set randomly. More specifically, both components of $\textbf{r}_c$ are set to be extracted from a uniform distribution on the interval $[-3,3]$, while the number of initial thermal excitations in the system is extracted from a uniform distribution on the interval $[0,5]$.

\section{Results}
\label{s:results}
In the following we will fix the unknown, but fixed, frequency, the squeezing rate and the efficiency of the homodyne measurement respectively to $\omega=0.1\kappa$, $\chi = 0.49\kappa$ and $\eta=0.9$, with $\kappa$ the cavity loss rate (see Eq. (\ref{eq:photocurrent1})). Our simulations show, however, that the agent is able to devise optimized feedback strategies in different regimes; in Appendix \ref{a:efficiency} we exemplify such flexibility of the proposed method by showing the results obtained for different values of the monitoring efficiency $\eta$. 
\par
We will denote our figure of merit, that is the effective QFI in Eq. (\ref{eq:effQFI}), as $\mathcal{Q}_{\sf eff}^{\sf(RL)}$. We will compare and contrast the results obtained by means of RL to  two benchmark strategies: the one where {\em no control} is applied, quantified by the figure of merit $\mathcal{Q}_{\sf eff}^{\sf (0)}$, and the strategy where, thanks to some a-priori information on the parameter $\omega$ (a typical assumption in the context of local quantum estimation theory \cite{MatteoIJQI}), a deterministic value of the control frequency is fixed as $\omega_{\sf fb} = - \omega$. Notice that in this latter case the control is deterministic and thus it does not correspond to a feedback, but rather to an {\em open-loop} (OL) control strategy, yielding the largest amount of conditional squeezing along the quadrature $\hat{q}$. The continuous monitoring, on the other hand, will yield a non-zero (but typically small) stochastic contribution on the $\hat{q}$ axis of phase-space. As a consequence the directions of squeezing and first moments will not be optimized. We will denote the figure of merit for this open-loop control strategy as $\mathcal{Q}_{\sf eff}^{\sf (OL)}$.

\begin{figure}
    \centering
    \includegraphics{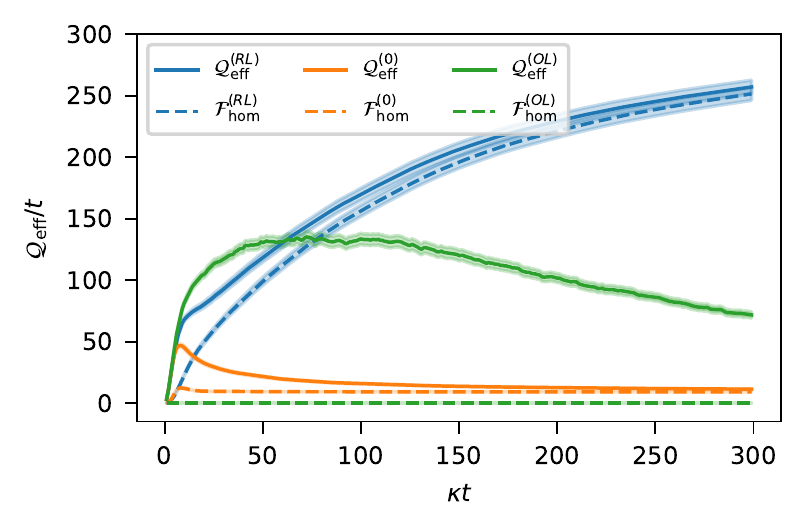}
    \caption{
    We compare the performances of the feedback strategy devised by the neural network with the two benchmark strategies described in the main text as a function of time. Solid lines correspond to the three different effective QFI ($\mathcal{Q}_{\sf eff}^{\sf (RL)}$,$\mathcal{Q}_{\sf eff}^{\sf (0)}$,$\mathcal{Q}_{\sf eff}^{\sf (OL)}$) divided by time, while dashed correspond to the the continuous monitoring classical FIs: ($\mathcal{F}_{\sf hom}^{\sf (RL)}$,$\mathcal{F}_{\sf hom}^{\sf (0)}$,$\mathcal{F}_{\sf hom}^{\sf (OL)}$), divided by time. The average QFI of the conditional states ($\mathcal{\bar{Q}}_c^{\sf (RL)}$,$\mathcal{\bar{Q}}_c^{\sf (0)}$,$\mathcal{\bar{Q}}_c^{\sf (OL)}$) can be derived as the difference between the two curves above.\\
    The results have been obtained simulating $N=5000$ trajectories with a time-step $dt=0.001/\kappa$, by fixing the parameters: $\omega=0.1\kappa$, $\chi=0.49\kappa$, $\eta=0.9$ and by considering as an initial state a thermal state with $n_{\sf th} = 5$ and initial first moment vector $\bar{\bf r}_c(0) = (0,0)$.}
    \label{f:compareQeff}
\end{figure}

The main result of this work is presented in Fig. \ref{f:compareQeff}: We considered as initial state a thermal state with $n_{\sf th}=5$ thermal excitations, and a first moment vector $\bar{\bf r}_c=(0,0)$. We show that, apart from an initial transient time where $\mathcal{Q}_{\sf eff}^{\sf(RL)} \lesssim \mathcal{Q}_{\sf eff}^{\sf(OL)}$, the feedback protocol yields a much larger effective QFI than the benchmark strategies considered. 
In particular, by looking at the behaviour of the two terms entering in Eq. (\ref{eq:effQFI}), we can make two main observations: i) as regards the average QFIs of the conditional states, that in Fig. \ref{f:compareQeff} correspond to the difference between the curves with the same colours, one finds $\bar{\mathcal{Q}}_c^{\sf (0)} < \bar{\mathcal{Q}}_c^{\sf (RL)} < \bar{\mathcal{Q}}_c^{\sf (OL)}$, that is, the feedback protocol is able to generate conditional states that are on average more sensitive respect to the one generated without feedback, but much less sensitive to the ones generated via the open-loop control protocol; ii) the enhancement in the estimation is thus mainly obtained thanks to the information contained in the continuous measurement outcomes: the monitoring FI $\mathcal{F}_{\sf hom}^{\sf (RL)}$ greatly overcomes the values of the same figure of merit for the two other protocols; iii) we observe that while for the open-loop control protocol $\bar{\mathcal{Q}}_c^{\sf (OL)}$ saturates to a given value once $\boldsymbol{\sigma}_c$ has reached its deterministic steady-state, the RL agent seems able to keep $\mathcal{F}_{\sf hom}^{\sf (RL)}$ increasing steadily in time, yielding a large enhancement in the long time limit. \\

Moreover, as in Eq. (\ref{eq:QCRB}) we observe that the relevant figure of merit in frequency estimation is $\mathcal{Q}_{\sf eff}/t$, if one allows to optimize over the single experiment monitoring time $t$ at fixed total time of the experiment $T$, the strategy devised by the agent clearly gives the best result. Our results hint also to the fact that in this case the optimization is obtained in the long-time limit, where the whole information is basically completely contained in the continuous homodyne measurement outcomes and the strong measurement on the conditional states is almost irrelevant (we however refer the reader to Appendix \ref{a:finalhomodyne} for a discussion on the effectiveness of homodyne detection as a final strong measurement for the three strategies considered). \\

\begin{figure}
    \centering
    \includegraphics{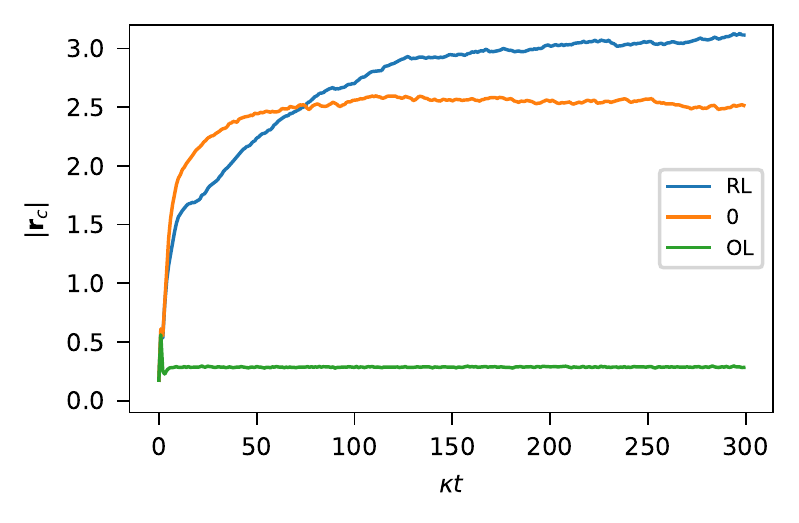}
    \caption{We plot $\mathbbm{E}_{\sf traj}[|\bar{\bf r}_c|]$ as a function of time for the three protocols considered (from top to bottom: RL-feedback, no-control, open-loop control). \\
  The results have been obtained simulating $N=5000$ trajectories with a time-step $dt=0.001/\kappa$, by fixing the parameters: $\omega=0.1$, $\chi=0.49\kappa$, $\eta=0.9$ and by considering as an initial state a thermal state with $n_{\sf th} = 5$ and initial first moment vector $\bar{\bf r}_c(0) = (0,0)$.}
    \label{f:absrc}
\end{figure}

In order to better understand these results, it is useful to look at the evolution of single trajectories and thus at the properties of the conditional states. As mentioned before, the achievable estimation precision will depend on the amount of squeezing generated during the dynamics and on its interplay with the first-moment vector $\bar{\bf r}_c$. In Fig. \ref{f:absrc} we compare the values of the magnitude of the first moments averaged over the trajectories  $\mathbbm{E}[|\bar{\bf r}_c|]$ for the three protocols. We indeed observe that the RL agent yields the largest values of $\mathbbm{E}[|\bar{\bf r}_c|]$, while the OL-control protocol yields almost negligible first moments.\\
We stressed before how squeezing is the main resource for this kind of estimation. In this respect, we know that the maximum amount of squeezing is generated deterministically in the OL-control protocol, yielding at steady state with $\xi^{\sf (OL)}\approx 6.05$ dB of squeezing for the values we considered in these simulations. However we discussed before that this squeezing is always {\em parallel} to the corresponding vector $\bar{\mathbf{r}}_c^{\sf (OL)}$; despite this fact and thanks to the fact the first moments are close to zero, this protocol still yields large values of $\mathcal{Q}[\varrho_c^{\sf (OL)}]$, as we indeed observed in Fig. \ref{f:compareQeff}. 

\begin{figure}
\centering
    \includegraphics[width=.95\columnwidth]{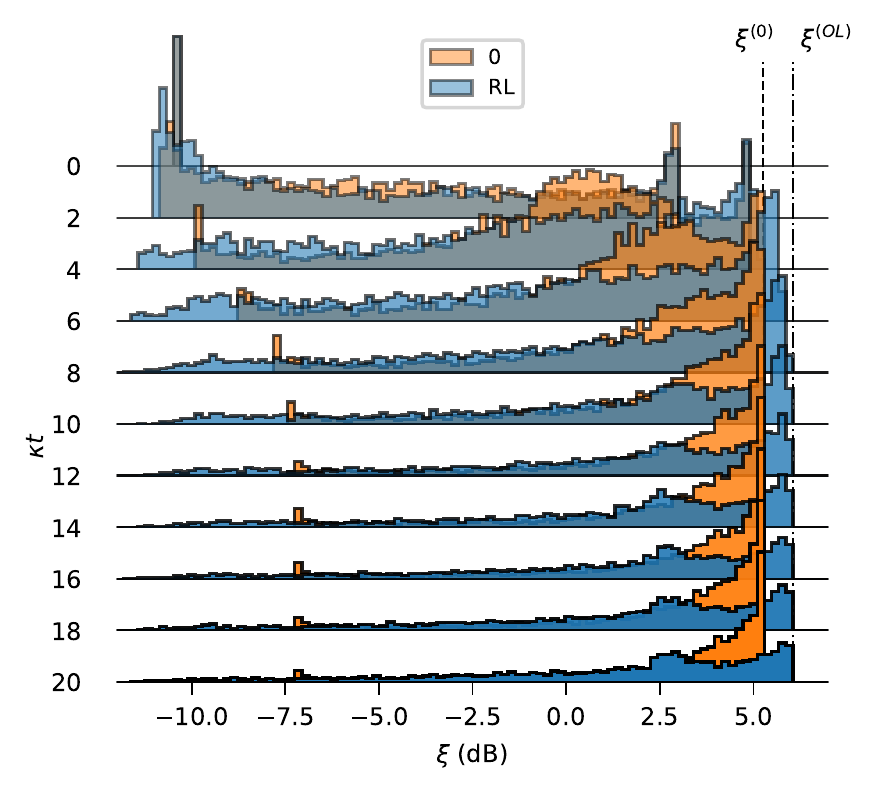}
    \caption{
    Histograms of the probability density of the squeezing $\xi$ (expressed in dB) perpendicular to the conditional first moment vector $\bar{\bf r}_c$ for different times and for both the no-control protocol (orange histograms) and for the RL agent feedback protocol (blue histograms). The two vertical lines correspond to the two bounds on the amount of squeezing for the no-control protocol $\xi^{(0)}$ and for the open-loop control protocol $\xi^{\sf (OL)}$.\\
    The results have been obtained simulating $N=5000$ trajectories with a time-step $dt=0.001/\kappa$, by fixing the parameters: $\omega=0.1$, $\chi=0.49\kappa$, $\eta=0.9$ and by considering as an initial state a thermal state with $n_{\sf th} = 5$ and initial first moment vector $\bar{\bf r}_c(0) = (0,0)$.}
    \label{f:perpsqueezing}
\end{figure}

If we now focus on the squeezing along the quadrature {\em perpendicular} to $\bar{\bf r}_c$ and thus possibly enhancing the contribution due to non-zero first moments in phase-space, we find non-trivial and definitely interesting results as shown in Fig. \ref{f:perpsqueezing}. 
In this figure we plot the histograms corresponding to the probability density of squeezing perpendicular to the conditional first moment vector $\bar{\bf r}_c$ for the no-control protocol and for the RL agent based feedback protocol for different times. 
As regards the protocol without control, we know that at steady-state one obtains a deterministic squeezing of $\xi^{(0)} \approx 5.25$dB, and as a consequence the squeezing along the quadrature perpendicular to the, stochastically varying, $\bar{\bf r}_c$ is going to be bounded by this value. This behaviour is indeed confirmed by looking at the orange histograms. If we now finally focus on the results corresponding to the RL agent based feedback (blue histograms), we can clearly observe how it is indeed able also to generate a large fraction of trajectories with squeezing perpendicular to $\bar{\bf r}_c$ not only well beyond the maximum value obtainable without control $\xi^{(0)}$, but also near to the limit $\xi^{\sf (OL)}$ achieved by the open-loop control discussed before 
and that we remind here is however always parallel to the first moment vector $\bar{\mathbf{r}}_c$. In particular we observe that, not only the RL-strategy is able to generate conditional states with the maximum squeezing achievable and with the most useful direction, but also the mode of this perpendicular squeezing distribution quickly saturates towards this limit $\xi^{\sf (OL)}$. 
Our results thus suggest how the portion of trajectories characterized by large first moments and large perpendicular squeezing is responsible for the enhancement in the frequency estimation precision. We refer to Appendices \ref{a:coupling} and \ref{a:efficiency} for some extra results that we have obtained by considering  different values of the coupling constant $\chi$ and of the monitoring efficiency $\eta$, and that further confirm our intuitions. For example when one considers smaller values of $\chi$, and as a consequence smaller amount of squeezing generated, all the the strategies considered yield as expected smaller values of the effective QFI. Similarly, we show how for smaller values of $\eta$, the effect on the squeezing generation is slightly reduced and that the main contribution to the enhancement is given by the first moment vector amplitude $|\bar{\bf r}_c|$.

\begin{figure}
    \centering 
    \includegraphics[width=0.9\columnwidth]{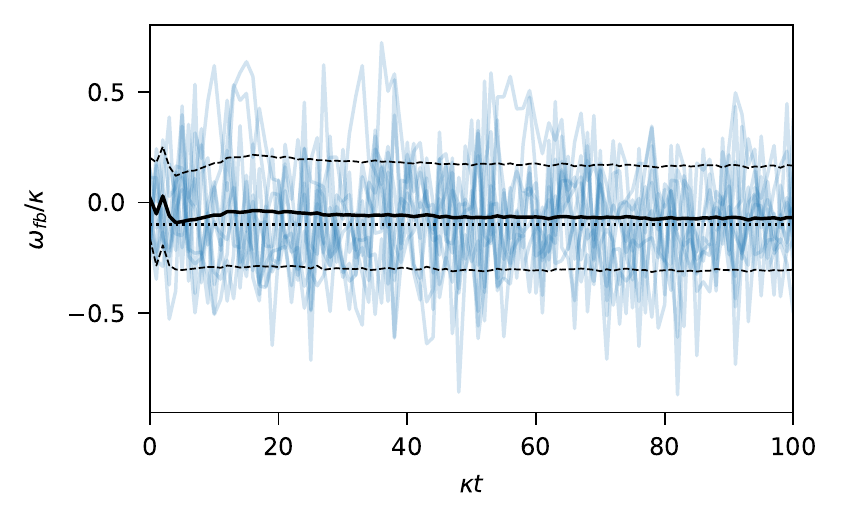}
    \caption{Feedback parameter $\omega_{\sf fb}$ obtained via the RL agent as a function of time, both for few samples trajectories (corresponding to blue lines with different shades of blue) and averaged over 5000 trajectories (black line). The dashed black lines show the standard deviation. The average of the RL is close to the value $-\omega$, yielding the maximum squeezing (dotted black line). The other physical parameters are set as in the previous figures.}
    \label{f:omegafb}
\end{figure}

It is also interesting to observe the behaviour of the feedback parameter $\omega_{\sf fb}$ as a function of time, both for a sample trajectory and averaged over the different trajectories. In Fig. \ref{f:omegafb} we find that the average value seems to converge to a value near to $\mathbbm{E}[\omega_{\sf fb}]\approx -\omega$, that is the one implemented in the open-loop control and yielding the maximum squeezing. However at the trajectory level the fluctuations of $\omega_{\sf fb}$ are evident and are thus crucial to increase $|\bar{\bf r}_c|$ and to optimize both the squeezing magnitude and more importantly its direction.

\par
Plainly speaking we can conclude that the feedback devised by the neural network is able to optimize the non-trivial interplay between first moments and squeezing and indeed to generate a significant amount of trajectories with larger amount of {\em perpendicular squeezing}. These trajectories are thus responsible for the enhancement in the estimation precision observed Fig. \ref{f:compareQeff}. Our results show also that this feature is much more relevant for the homodyne FI $\mathcal{F}_{\sf hom}$, that is indeed responsible to the enhancement yielded by the feedback strategy. A hint in this direction is already given by the formula (\ref{eq:Fhom}) for $\mathcal{F}_{\sf hom}$, that depends directly on the vector $\partial_{\omega}\bar{\bf r}_c$ (however we should remark that the evolution of $\partial_{\omega}\bar{\bf r}_c$ in Eq. (\ref{eq:drc})  depends also on $\boldsymbol{\sigma}_c$ and thus on the squeezing properties of the conditional states).\\

\section{Conclusions}
\label{s:conclusion}
In this work we have shown how a RL algorithm has been able to optimize a feedback strategy able to attain a high precision in frequency estimation. We have understood the results in terms of the optimization of the interplay between the amplitude and the squeezing generated by the protocol. This kind of optimization is highly non-trivial: a simple strategy trying to optimize this kind of feature at each time $t$ cannot be devised because of the stochasticity of the subsequent evolution.
\par
As a future outlook we aim to optimize the neural network, in order to be able to reduce the needed observations ${\bf obs}$. In particular we will look at strategies able to exploit just the real-time measurement output ${\bf dy}_t$, and thus corresponding to Markovian feedback~\cite{Wiseman1993,Wiseman1994}
\par
We have witnessed a great experimental improvement in the implementation of FPGA-based real-time state-based feedback, as shown recently in the context of the cooling of mechanical oscillators~\cite{Rossi2018,Magrini2021,Tebbenjohanns2021}. We remark that once a neural network has been trained, its real-time interrogation is not much more computationally costly than what has been done in the cited experiments. We are thus confident that feedback strategies previously trained via RL algorithms can be efficently implemented in the next future for quantum metrology purposes as we have described, or for more general quantum technological tasks.
\begin{acknowledgments}
We thank F. Albarelli and M. Paris for helpful discussions.
MACR acknowledges financial support from the Academy of Finland via the Centre of Excellence program (Project no. 336810).
MGG and DT acknowledge support from the Sviluppo UniMi 2018 initiative.
The computer resources of the Finnish IT Center for Science (CSC) and the FGCI project (Finland) are acknowledged.
\end{acknowledgments}
\bibliography{library} 

\begin{thebibliography}{76}%
\makeatletter
\providecommand \@ifxundefined [1]{%
 \@ifx{#1\undefined}
}%
\providecommand \@ifnum [1]{%
 \ifnum #1\expandafter \@firstoftwo
 \else \expandafter \@secondoftwo
 \fi
}%
\providecommand \@ifx [1]{%
 \ifx #1\expandafter \@firstoftwo
 \else \expandafter \@secondoftwo
 \fi
}%
\providecommand \natexlab [1]{#1}%
\providecommand \enquote  [1]{``#1''}%
\providecommand \bibnamefont  [1]{#1}%
\providecommand \bibfnamefont [1]{#1}%
\providecommand \citenamefont [1]{#1}%
\providecommand \href@noop [0]{\@secondoftwo}%
\providecommand \href [0]{\begingroup \@sanitize@url \@href}%
\providecommand \@href[1]{\@@startlink{#1}\@@href}%
\providecommand \@@href[1]{\endgroup#1\@@endlink}%
\providecommand \@sanitize@url [0]{\catcode `\\12\catcode `\$12\catcode
  `\&12\catcode `\#12\catcode `\^12\catcode `\_12\catcode `\%12\relax}%
\providecommand \@@startlink[1]{}%
\providecommand \@@endlink[0]{}%
\providecommand \url  [0]{\begingroup\@sanitize@url \@url }%
\providecommand \@url [1]{\endgroup\@href {#1}{\urlprefix }}%
\providecommand \urlprefix  [0]{URL }%
\providecommand \Eprint [0]{\href }%
\providecommand \doibase [0]{http://dx.doi.org/}%
\providecommand \selectlanguage [0]{\@gobble}%
\providecommand \bibinfo  [0]{\@secondoftwo}%
\providecommand \bibfield  [0]{\@secondoftwo}%
\providecommand \translation [1]{[#1]}%
\providecommand \BibitemOpen [0]{}%
\providecommand \bibitemStop [0]{}%
\providecommand \bibitemNoStop [0]{.\EOS\space}%
\providecommand \EOS [0]{\spacefactor3000\relax}%
\providecommand \BibitemShut  [1]{\csname bibitem#1\endcsname}%
\let\auto@bib@innerbib\@empty
\bibitem [{\citenamefont {Giovannetti}\ \emph {et~al.}(2011)\citenamefont
  {Giovannetti}, \citenamefont {Lloyd},\ and\ \citenamefont
  {Maccone}}]{GiovannettiNatPhot}%
  \BibitemOpen
  \bibfield  {author} {\bibinfo {author} {\bibfnamefont {V.}~\bibnamefont
  {Giovannetti}}, \bibinfo {author} {\bibfnamefont {S.}~\bibnamefont {Lloyd}},
  \ and\ \bibinfo {author} {\bibfnamefont {L.}~\bibnamefont {Maccone}},\ }\href
  {\doibase 10.1038/nphoton.2011.35} {\bibfield  {journal} {\bibinfo  {journal}
  {Nat. Photonics}\ }\textbf {\bibinfo {volume} {5}},\ \bibinfo {pages} {222}
  (\bibinfo {year} {2011})},\ \Eprint {http://arxiv.org/abs/1102.2318}
  {1102.2318} \BibitemShut {NoStop}%
\bibitem [{\citenamefont {Pirandola}\ \emph {et~al.}(2018)\citenamefont
  {Pirandola}, \citenamefont {Bardhan}, \citenamefont {Gehring}, \citenamefont
  {Weedbrook},\ and\ \citenamefont {Lloyd}}]{PirandolaReview}%
  \BibitemOpen
  \bibfield  {author} {\bibinfo {author} {\bibfnamefont {S.}~\bibnamefont
  {Pirandola}}, \bibinfo {author} {\bibfnamefont {B.~R.}\ \bibnamefont
  {Bardhan}}, \bibinfo {author} {\bibfnamefont {T.}~\bibnamefont {Gehring}},
  \bibinfo {author} {\bibfnamefont {C.}~\bibnamefont {Weedbrook}}, \ and\
  \bibinfo {author} {\bibfnamefont {S.}~\bibnamefont {Lloyd}},\ }\href
  {\doibase 10.1038/s41566-018-0301-6} {\bibfield  {journal} {\bibinfo
  {journal} {Nature Photonics}\ }\textbf {\bibinfo {volume} {12}},\ \bibinfo
  {pages} {724} (\bibinfo {year} {2018})}\BibitemShut {NoStop}%
\bibitem [{\citenamefont {Wiseman}\ and\ \citenamefont
  {Milburn}(2010)}]{WisemanMilburn}%
  \BibitemOpen
  \bibfield  {author} {\bibinfo {author} {\bibfnamefont {H.~M.}\ \bibnamefont
  {Wiseman}}\ and\ \bibinfo {author} {\bibfnamefont {G.~J.}\ \bibnamefont
  {Milburn}},\ }\href@noop {} {\emph {\bibinfo {title} {{Quantum Measurement
  and Control}}}}\ (\bibinfo  {publisher} {Cambridge University Press},\
  \bibinfo {address} {New York},\ \bibinfo {year} {2010})\BibitemShut {NoStop}%
\bibitem [{\citenamefont {Jacobs}\ and\ \citenamefont
  {Steck}(2006)}]{SteckJacobs}%
  \BibitemOpen
  \bibfield  {author} {\bibinfo {author} {\bibfnamefont {K.}~\bibnamefont
  {Jacobs}}\ and\ \bibinfo {author} {\bibfnamefont {D.~A.}\ \bibnamefont
  {Steck}},\ }\href {\doibase 10.1080/00107510601101934} {\bibfield  {journal}
  {\bibinfo  {journal} {Contemp. Phys.}\ }\textbf {\bibinfo {volume} {47}},\
  \bibinfo {pages} {279} (\bibinfo {year} {2006})},\ \Eprint
  {http://arxiv.org/abs/quant-ph/0611067} {quant-ph/0611067} \BibitemShut
  {NoStop}%
\bibitem [{\citenamefont {Guţă}\ \emph {et~al.}(2007)\citenamefont {Guţă},
  \citenamefont {Janssens},\ and\ \citenamefont {Kahn}}]{Guta2007a}%
  \BibitemOpen
  \bibfield  {author} {\bibinfo {author} {\bibfnamefont {M.}~\bibnamefont
  {Guţă}}, \bibinfo {author} {\bibfnamefont {B.}~\bibnamefont {Janssens}}, \
  and\ \bibinfo {author} {\bibfnamefont {J.}~\bibnamefont {Kahn}},\ }\href
  {\doibase 10.1007/s00220-007-0357-5} {\bibfield  {journal} {\bibinfo
  {journal} {Commun. Math. Phys.}\ }\textbf {\bibinfo {volume} {277}},\
  \bibinfo {pages} {127} (\bibinfo {year} {2007})},\ \Eprint
  {http://arxiv.org/abs/0608074v3} {arXiv:0608074v3 [arXiv:quant-ph]}
  \BibitemShut {NoStop}%
\bibitem [{\citenamefont {Tsang}\ \emph {et~al.}(2011)\citenamefont {Tsang},
  \citenamefont {Wiseman},\ and\ \citenamefont {Caves}}]{Tsang2011}%
  \BibitemOpen
  \bibfield  {author} {\bibinfo {author} {\bibfnamefont {M.}~\bibnamefont
  {Tsang}}, \bibinfo {author} {\bibfnamefont {H.~M.}\ \bibnamefont {Wiseman}},
  \ and\ \bibinfo {author} {\bibfnamefont {C.~M.}\ \bibnamefont {Caves}},\
  }\href {\doibase 10.1103/PhysRevLett.106.090401} {\bibfield  {journal}
  {\bibinfo  {journal} {Phys. Rev. Lett.}\ }\textbf {\bibinfo {volume} {106}},\
  \bibinfo {pages} {090401} (\bibinfo {year} {2011})}\BibitemShut {NoStop}%
\bibitem [{\citenamefont {Tsang}(2013)}]{Tsang2013a}%
  \BibitemOpen
  \bibfield  {author} {\bibinfo {author} {\bibfnamefont {M.}~\bibnamefont
  {Tsang}},\ }\href {\doibase 10.1088/1367-2630/15/7/073005} {\bibfield
  {journal} {\bibinfo  {journal} {New J. Phys.}\ }\textbf {\bibinfo {volume}
  {15}},\ \bibinfo {pages} {73005} (\bibinfo {year} {2013})},\ \Eprint
  {http://arxiv.org/abs/1301.5733} {1301.5733} \BibitemShut {NoStop}%
\bibitem [{\citenamefont {Gammelmark}\ and\ \citenamefont
  {M{\o}lmer}(2013{\natexlab{a}})}]{GammelmarkCRB}%
  \BibitemOpen
  \bibfield  {author} {\bibinfo {author} {\bibfnamefont {S.}~\bibnamefont
  {Gammelmark}}\ and\ \bibinfo {author} {\bibfnamefont {K.}~\bibnamefont
  {M{\o}lmer}},\ }\href {\doibase 10.1103/PhysRevA.87.032115} {\bibfield
  {journal} {\bibinfo  {journal} {Phys. Rev. A}\ }\textbf {\bibinfo {volume}
  {87}},\ \bibinfo {pages} {032115} (\bibinfo {year} {2013}{\natexlab{a}})},\
  \Eprint {http://arxiv.org/abs/1212.5700} {1212.5700} \BibitemShut {NoStop}%
\bibitem [{\citenamefont {Gammelmark}\ and\ \citenamefont
  {M{\o}lmer}(2014)}]{GammelmarkQCRB}%
  \BibitemOpen
  \bibfield  {author} {\bibinfo {author} {\bibfnamefont {S.}~\bibnamefont
  {Gammelmark}}\ and\ \bibinfo {author} {\bibfnamefont {K.}~\bibnamefont
  {M{\o}lmer}},\ }\href {\doibase 10.1103/PhysRevLett.112.170401} {\bibfield
  {journal} {\bibinfo  {journal} {Phys. Rev. Lett.}\ }\textbf {\bibinfo
  {volume} {112}},\ \bibinfo {pages} {170401} (\bibinfo {year}
  {2014})}\BibitemShut {NoStop}%
\bibitem [{\citenamefont {Guţă}\ and\ \citenamefont
  {Kiukas}(2017)}]{Guta2016}%
  \BibitemOpen
  \bibfield  {author} {\bibinfo {author} {\bibfnamefont {M.}~\bibnamefont
  {Guţă}}\ and\ \bibinfo {author} {\bibfnamefont {J.}~\bibnamefont
  {Kiukas}},\ }\href {\doibase 10.1063/1.4982958} {\bibfield  {journal}
  {\bibinfo  {journal} {J. Mat. Phys.}\ }\textbf {\bibinfo {volume} {58}},\
  \bibinfo {pages} {052201} (\bibinfo {year} {2017})}\BibitemShut {NoStop}%
\bibitem [{\citenamefont {Genoni}(2017)}]{Genoni2017}%
  \BibitemOpen
  \bibfield  {author} {\bibinfo {author} {\bibfnamefont {M.~G.}\ \bibnamefont
  {Genoni}},\ }\href {\doibase 10.1103/PhysRevA.95.012116} {\bibfield
  {journal} {\bibinfo  {journal} {Phys. Rev. A}\ }\textbf {\bibinfo {volume}
  {95}},\ \bibinfo {pages} {012116} (\bibinfo {year} {2017})},\ \Eprint
  {http://arxiv.org/abs/1608.08429} {1608.08429} \BibitemShut {NoStop}%
\bibitem [{\citenamefont {Albarelli}\ \emph {et~al.}(2017)\citenamefont
  {Albarelli}, \citenamefont {Rossi}, \citenamefont {Paris},\ and\
  \citenamefont {Genoni}}]{Albarelli2017a}%
  \BibitemOpen
  \bibfield  {author} {\bibinfo {author} {\bibfnamefont {F.}~\bibnamefont
  {Albarelli}}, \bibinfo {author} {\bibfnamefont {M.~A.~C.}\ \bibnamefont
  {Rossi}}, \bibinfo {author} {\bibfnamefont {M.~G.~A.}\ \bibnamefont {Paris}},
  \ and\ \bibinfo {author} {\bibfnamefont {M.~G.}\ \bibnamefont {Genoni}},\
  }\href {\doibase 10.1088/1367-2630/aa9840} {\bibfield  {journal} {\bibinfo
  {journal} {New J. Phys.}\ }\textbf {\bibinfo {volume} {19}},\ \bibinfo
  {pages} {123011} (\bibinfo {year} {2017})},\ \Eprint
  {http://arxiv.org/abs/1706.00485} {1706.00485} \BibitemShut {NoStop}%
\bibitem [{\citenamefont {Mabuchi}(1996)}]{Mabuchi1996}%
  \BibitemOpen
  \bibfield  {author} {\bibinfo {author} {\bibfnamefont {H.}~\bibnamefont
  {Mabuchi}},\ }\href {http://stacks.iop.org/1355-5111/8/i=6/a=002} {\bibfield
  {journal} {\bibinfo  {journal} {Quant. Semiclass. Opt.}\ }\textbf {\bibinfo
  {volume} {8}},\ \bibinfo {pages} {1103} (\bibinfo {year} {1996})}\BibitemShut
  {NoStop}%
\bibitem [{\citenamefont {Gambetta}\ and\ \citenamefont
  {Wiseman}(2001)}]{Gambetta2001}%
  \BibitemOpen
  \bibfield  {author} {\bibinfo {author} {\bibfnamefont {J.}~\bibnamefont
  {Gambetta}}\ and\ \bibinfo {author} {\bibfnamefont {H.~M.}\ \bibnamefont
  {Wiseman}},\ }\href {\doibase 10.1103/PhysRevA.64.042105} {\bibfield
  {journal} {\bibinfo  {journal} {Phys. Rev. A}\ }\textbf {\bibinfo {volume}
  {64}},\ \bibinfo {pages} {042105} (\bibinfo {year} {2001})}\BibitemShut
  {NoStop}%
\bibitem [{\citenamefont {Geremia}\ \emph {et~al.}(2003)\citenamefont
  {Geremia}, \citenamefont {Stockton}, \citenamefont {Doherty},\ and\
  \citenamefont {Mabuchi}}]{Geremia2003}%
  \BibitemOpen
  \bibfield  {author} {\bibinfo {author} {\bibfnamefont {J.~M.}\ \bibnamefont
  {Geremia}}, \bibinfo {author} {\bibfnamefont {J.~K.}\ \bibnamefont
  {Stockton}}, \bibinfo {author} {\bibfnamefont {A.~C.}\ \bibnamefont
  {Doherty}}, \ and\ \bibinfo {author} {\bibfnamefont {H.}~\bibnamefont
  {Mabuchi}},\ }\href {\doibase 10.1103/PhysRevLett.91.250801} {\bibfield
  {journal} {\bibinfo  {journal} {Phys. Rev. Lett.}\ }\textbf {\bibinfo
  {volume} {91}},\ \bibinfo {pages} {250801} (\bibinfo {year}
  {2003})}\BibitemShut {NoStop}%
\bibitem [{\citenamefont {M{\o}lmer}\ and\ \citenamefont
  {Madsen}(2004)}]{Molmer2004}%
  \BibitemOpen
  \bibfield  {author} {\bibinfo {author} {\bibfnamefont {K.}~\bibnamefont
  {M{\o}lmer}}\ and\ \bibinfo {author} {\bibfnamefont {L.~B.}\ \bibnamefont
  {Madsen}},\ }\href {\doibase 10.1103/PhysRevA.70.052102} {\bibfield
  {journal} {\bibinfo  {journal} {Phys. Rev. A}\ }\textbf {\bibinfo {volume}
  {70}},\ \bibinfo {pages} {052102} (\bibinfo {year} {2004})},\ \Eprint
  {http://arxiv.org/abs/quant-ph/0402158} {quant-ph/0402158} \BibitemShut
  {NoStop}%
\bibitem [{\citenamefont {Madsen}\ and\ \citenamefont
  {M{\o}lmer}(2004)}]{Madsen2004}%
  \BibitemOpen
  \bibfield  {author} {\bibinfo {author} {\bibfnamefont {L.~B.}\ \bibnamefont
  {Madsen}}\ and\ \bibinfo {author} {\bibfnamefont {K.}~\bibnamefont
  {M{\o}lmer}},\ }\href {\doibase 10.1103/PhysRevA.70.052324} {\bibfield
  {journal} {\bibinfo  {journal} {Phys. Rev. A}\ }\textbf {\bibinfo {volume}
  {70}},\ \bibinfo {pages} {052324} (\bibinfo {year} {2004})}\BibitemShut
  {NoStop}%
\bibitem [{\citenamefont {Stockton}\ \emph {et~al.}(2004)\citenamefont
  {Stockton}, \citenamefont {Geremia}, \citenamefont {Doherty},\ and\
  \citenamefont {Mabuchi}}]{Stockton2004}%
  \BibitemOpen
  \bibfield  {author} {\bibinfo {author} {\bibfnamefont {J.~K.}\ \bibnamefont
  {Stockton}}, \bibinfo {author} {\bibfnamefont {J.~M.}\ \bibnamefont
  {Geremia}}, \bibinfo {author} {\bibfnamefont {A.~C.}\ \bibnamefont
  {Doherty}}, \ and\ \bibinfo {author} {\bibfnamefont {H.}~\bibnamefont
  {Mabuchi}},\ }\href {\doibase 10.1103/PhysRevA.69.032109} {\bibfield
  {journal} {\bibinfo  {journal} {Phys. Rev. A}\ }\textbf {\bibinfo {volume}
  {69}},\ \bibinfo {pages} {032109} (\bibinfo {year} {2004})},\ \Eprint
  {http://arxiv.org/abs/quant-ph/0309101} {quant-ph/0309101} \BibitemShut
  {NoStop}%
\bibitem [{\citenamefont {Tsang}(2010)}]{Tsang2010}%
  \BibitemOpen
  \bibfield  {author} {\bibinfo {author} {\bibfnamefont {M.}~\bibnamefont
  {Tsang}},\ }\href {\doibase 10.1103/PhysRevA.81.013824} {\bibfield  {journal}
  {\bibinfo  {journal} {Phys. Rev. A}\ }\textbf {\bibinfo {volume} {81}},\
  \bibinfo {pages} {013824} (\bibinfo {year} {2010})},\ \Eprint
  {http://arxiv.org/abs/0909.2432} {arXiv:0909.2432} \BibitemShut {NoStop}%
\bibitem [{\citenamefont {Wheatley}\ \emph {et~al.}(2010)\citenamefont
  {Wheatley}, \citenamefont {Berry}, \citenamefont {Yonezawa}, \citenamefont
  {Nakane}, \citenamefont {Arao}, \citenamefont {Pope}, \citenamefont {Ralph},
  \citenamefont {Wiseman}, \citenamefont {Furusawa},\ and\ \citenamefont
  {Huntington}}]{Wheatley2010}%
  \BibitemOpen
  \bibfield  {author} {\bibinfo {author} {\bibfnamefont {T.~A.}\ \bibnamefont
  {Wheatley}}, \bibinfo {author} {\bibfnamefont {D.~W.}\ \bibnamefont {Berry}},
  \bibinfo {author} {\bibfnamefont {H.}~\bibnamefont {Yonezawa}}, \bibinfo
  {author} {\bibfnamefont {D.}~\bibnamefont {Nakane}}, \bibinfo {author}
  {\bibfnamefont {H.}~\bibnamefont {Arao}}, \bibinfo {author} {\bibfnamefont
  {D.~T.}\ \bibnamefont {Pope}}, \bibinfo {author} {\bibfnamefont {T.~C.}\
  \bibnamefont {Ralph}}, \bibinfo {author} {\bibfnamefont {H.~M.}\ \bibnamefont
  {Wiseman}}, \bibinfo {author} {\bibfnamefont {A.}~\bibnamefont {Furusawa}}, \
  and\ \bibinfo {author} {\bibfnamefont {E.~H.}\ \bibnamefont {Huntington}},\
  }\href {\doibase 10.1103/PhysRevLett.104.093601} {\bibfield  {journal}
  {\bibinfo  {journal} {Phys. Rev. Lett.}\ }\textbf {\bibinfo {volume} {104}},\
  \bibinfo {pages} {093601} (\bibinfo {year} {2010})}\BibitemShut {NoStop}%
\bibitem [{\citenamefont {Yonezawa}\ \emph {et~al.}(2012)\citenamefont
  {Yonezawa}, \citenamefont {Nakane}, \citenamefont {Wheatley}, \citenamefont
  {Iwasawa}, \citenamefont {Takeda}, \citenamefont {Arao}, \citenamefont
  {Ohki}, \citenamefont {Tsumura}, \citenamefont {Berry}, \citenamefont
  {Ralph}, \citenamefont {Wiseman}, \citenamefont {Huntington},\ and\
  \citenamefont {Furusawa}}]{Yonezawa2012}%
  \BibitemOpen
  \bibfield  {author} {\bibinfo {author} {\bibfnamefont {H.}~\bibnamefont
  {Yonezawa}}, \bibinfo {author} {\bibfnamefont {D.}~\bibnamefont {Nakane}},
  \bibinfo {author} {\bibfnamefont {T.~A.}\ \bibnamefont {Wheatley}}, \bibinfo
  {author} {\bibfnamefont {K.}~\bibnamefont {Iwasawa}}, \bibinfo {author}
  {\bibfnamefont {S.}~\bibnamefont {Takeda}}, \bibinfo {author} {\bibfnamefont
  {H.}~\bibnamefont {Arao}}, \bibinfo {author} {\bibfnamefont {K.}~\bibnamefont
  {Ohki}}, \bibinfo {author} {\bibfnamefont {K.}~\bibnamefont {Tsumura}},
  \bibinfo {author} {\bibfnamefont {D.~W.}\ \bibnamefont {Berry}}, \bibinfo
  {author} {\bibfnamefont {T.~C.}\ \bibnamefont {Ralph}}, \bibinfo {author}
  {\bibfnamefont {H.~M.}\ \bibnamefont {Wiseman}}, \bibinfo {author}
  {\bibfnamefont {E.~H.}\ \bibnamefont {Huntington}}, \ and\ \bibinfo {author}
  {\bibfnamefont {A.}~\bibnamefont {Furusawa}},\ }\href
  {https://doi.org/10.1126/science.1225258} {\bibfield  {journal} {\bibinfo
  {journal} {Science}\ }\textbf {\bibinfo {volume} {337}},\ \bibinfo {pages}
  {1514} (\bibinfo {year} {2012})}\BibitemShut {NoStop}%
\bibitem [{\citenamefont {Cook}\ \emph {et~al.}(2014)\citenamefont {Cook},
  \citenamefont {Riofr{\'{i}}o},\ and\ \citenamefont {Deutsch}}]{Cook2014}%
  \BibitemOpen
  \bibfield  {author} {\bibinfo {author} {\bibfnamefont {R.~L.}\ \bibnamefont
  {Cook}}, \bibinfo {author} {\bibfnamefont {C.~A.}\ \bibnamefont
  {Riofr{\'{i}}o}}, \ and\ \bibinfo {author} {\bibfnamefont {I.~H.}\
  \bibnamefont {Deutsch}},\ }\href {\doibase 10.1103/PhysRevA.90.032113}
  {\bibfield  {journal} {\bibinfo  {journal} {Phys. Rev. A}\ }\textbf {\bibinfo
  {volume} {90}},\ \bibinfo {pages} {032113} (\bibinfo {year} {2014})},\
  \Eprint {http://arxiv.org/abs/1406.4482} {arXiv:1406.4482} \BibitemShut
  {NoStop}%
\bibitem [{\citenamefont {Six}\ \emph {et~al.}(2015)\citenamefont {Six},
  \citenamefont {Campagne-Ibarcq}, \citenamefont {Bretheau}, \citenamefont
  {Huard},\ and\ \citenamefont {Rouchon}}]{Six2015}%
  \BibitemOpen
  \bibfield  {author} {\bibinfo {author} {\bibfnamefont {P.}~\bibnamefont
  {Six}}, \bibinfo {author} {\bibfnamefont {P.}~\bibnamefont
  {Campagne-Ibarcq}}, \bibinfo {author} {\bibfnamefont {L.}~\bibnamefont
  {Bretheau}}, \bibinfo {author} {\bibfnamefont {B.}~\bibnamefont {Huard}}, \
  and\ \bibinfo {author} {\bibfnamefont {P.}~\bibnamefont {Rouchon}},\ }in\
  \href {\doibase 10.1109/CDC.2015.7403443} {\emph {\bibinfo {booktitle} {2015
  54th IEEE Conf. Decis. Control}}},\ \bibinfo {series and number} {\bibinfo
  {number} {Cdc}}\ (\bibinfo  {publisher} {IEEE},\ \bibinfo {year} {2015})\ p.\
  \bibinfo {pages} {7742}\BibitemShut {NoStop}%
\bibitem [{\citenamefont {Kiilerich}\ and\ \citenamefont
  {M{\o}lmer}(2016)}]{KiilerichHomodyne}%
  \BibitemOpen
  \bibfield  {author} {\bibinfo {author} {\bibfnamefont {A.~H.}\ \bibnamefont
  {Kiilerich}}\ and\ \bibinfo {author} {\bibfnamefont {K.}~\bibnamefont
  {M{\o}lmer}},\ }\href {\doibase 10.1103/PhysRevA.94.032103} {\bibfield
  {journal} {\bibinfo  {journal} {Phys. Rev. A}\ }\textbf {\bibinfo {volume}
  {94}},\ \bibinfo {pages} {032103} (\bibinfo {year} {2016})}\BibitemShut
  {NoStop}%
\bibitem [{\citenamefont {Cortez}\ \emph {et~al.}(2017)\citenamefont {Cortez},
  \citenamefont {Chantasri}, \citenamefont {Garc{\'{i}}a-Pintos}, \citenamefont
  {Dressel},\ and\ \citenamefont {Jordan}}]{Cortez2017}%
  \BibitemOpen
  \bibfield  {author} {\bibinfo {author} {\bibfnamefont {L.}~\bibnamefont
  {Cortez}}, \bibinfo {author} {\bibfnamefont {A.}~\bibnamefont {Chantasri}},
  \bibinfo {author} {\bibfnamefont {L.~P.}\ \bibnamefont
  {Garc{\'{i}}a-Pintos}}, \bibinfo {author} {\bibfnamefont {J.}~\bibnamefont
  {Dressel}}, \ and\ \bibinfo {author} {\bibfnamefont {A.~N.}\ \bibnamefont
  {Jordan}},\ }\href {\doibase 10.1103/PhysRevA.95.012314} {\bibfield
  {journal} {\bibinfo  {journal} {Phys. Rev. A}\ }\textbf {\bibinfo {volume}
  {95}},\ \bibinfo {pages} {012314} (\bibinfo {year} {2017})},\ \Eprint
  {http://arxiv.org/abs/1606.01407} {1606.01407} \BibitemShut {NoStop}%
\bibitem [{\citenamefont {Ralph}\ \emph {et~al.}(2017)\citenamefont {Ralph},
  \citenamefont {Maskell},\ and\ \citenamefont {Jacobs}}]{Ralph2017}%
  \BibitemOpen
  \bibfield  {author} {\bibinfo {author} {\bibfnamefont {J.~F.}\ \bibnamefont
  {Ralph}}, \bibinfo {author} {\bibfnamefont {S.}~\bibnamefont {Maskell}}, \
  and\ \bibinfo {author} {\bibfnamefont {K.}~\bibnamefont {Jacobs}},\ }\href
  {\doibase 10.1103/PhysRevA.96.052306} {\bibfield  {journal} {\bibinfo
  {journal} {Phys. Rev. A}\ }\textbf {\bibinfo {volume} {96}},\ \bibinfo
  {pages} {052306} (\bibinfo {year} {2017})},\ \Eprint
  {http://arxiv.org/abs/1707.04725} {1707.04725} \BibitemShut {NoStop}%
\bibitem [{\citenamefont {Atalaya}\ \emph {et~al.}(2018)\citenamefont
  {Atalaya}, \citenamefont {Hacohen-Gourgy}, \citenamefont {Martin},
  \citenamefont {Siddiqi},\ and\ \citenamefont {Korotkov}}]{Atalaya2017}%
  \BibitemOpen
  \bibfield  {author} {\bibinfo {author} {\bibfnamefont {J.}~\bibnamefont
  {Atalaya}}, \bibinfo {author} {\bibfnamefont {S.}~\bibnamefont
  {Hacohen-Gourgy}}, \bibinfo {author} {\bibfnamefont {L.~S.}\ \bibnamefont
  {Martin}}, \bibinfo {author} {\bibfnamefont {I.}~\bibnamefont {Siddiqi}}, \
  and\ \bibinfo {author} {\bibfnamefont {A.~N.}\ \bibnamefont {Korotkov}},\
  }\href {\doibase 10.1038/s41534-018-0091-1} {\bibfield  {journal} {\bibinfo
  {journal} {npj Quantum Inf.}\ }\textbf {\bibinfo {volume} {4}},\ \bibinfo
  {pages} {41} (\bibinfo {year} {2018})},\ \Eprint
  {http://arxiv.org/abs/1702.08077} {arXiv:1702.08077} \BibitemShut {NoStop}%
\bibitem [{\citenamefont {Albarelli}\ \emph {et~al.}(2018)\citenamefont
  {Albarelli}, \citenamefont {Rossi}, \citenamefont {Tamascelli},\ and\
  \citenamefont {Genoni}}]{Albarelli2018Quantum}%
  \BibitemOpen
  \bibfield  {author} {\bibinfo {author} {\bibfnamefont {F.}~\bibnamefont
  {Albarelli}}, \bibinfo {author} {\bibfnamefont {M.~A.~C.}\ \bibnamefont
  {Rossi}}, \bibinfo {author} {\bibfnamefont {D.}~\bibnamefont {Tamascelli}}, \
  and\ \bibinfo {author} {\bibfnamefont {M.~G.}\ \bibnamefont {Genoni}},\
  }\href {\doibase 10.22331/q-2018-12-03-110} {\bibfield  {journal} {\bibinfo
  {journal} {{Quantum}}\ }\textbf {\bibinfo {volume} {2}},\ \bibinfo {pages}
  {110} (\bibinfo {year} {2018})}\BibitemShut {NoStop}%
\bibitem [{\citenamefont {Shankar}\ \emph {et~al.}(2019)\citenamefont
  {Shankar}, \citenamefont {Greve}, \citenamefont {Wu}, \citenamefont
  {Thompson},\ and\ \citenamefont {Holland}}]{Shankar2019}%
  \BibitemOpen
  \bibfield  {author} {\bibinfo {author} {\bibfnamefont {A.}~\bibnamefont
  {Shankar}}, \bibinfo {author} {\bibfnamefont {G.~P.}\ \bibnamefont {Greve}},
  \bibinfo {author} {\bibfnamefont {B.}~\bibnamefont {Wu}}, \bibinfo {author}
  {\bibfnamefont {J.~K.}\ \bibnamefont {Thompson}}, \ and\ \bibinfo {author}
  {\bibfnamefont {M.}~\bibnamefont {Holland}},\ }\href {\doibase
  10.1103/PhysRevLett.122.233602} {\bibfield  {journal} {\bibinfo  {journal}
  {Phys. Rev. Lett.}\ }\textbf {\bibinfo {volume} {122}},\ \bibinfo {pages}
  {233602} (\bibinfo {year} {2019})}\BibitemShut {NoStop}%
\bibitem [{\citenamefont {Rossi}\ \emph {et~al.}(2020)\citenamefont {Rossi},
  \citenamefont {Albarelli}, \citenamefont {Tamascelli},\ and\ \citenamefont
  {Genoni}}]{Rossi2020PRL}%
  \BibitemOpen
  \bibfield  {author} {\bibinfo {author} {\bibfnamefont {M.~A.~C.}\
  \bibnamefont {Rossi}}, \bibinfo {author} {\bibfnamefont {F.}~\bibnamefont
  {Albarelli}}, \bibinfo {author} {\bibfnamefont {D.}~\bibnamefont
  {Tamascelli}}, \ and\ \bibinfo {author} {\bibfnamefont {M.~G.}\ \bibnamefont
  {Genoni}},\ }\href {\doibase 10.1103/PhysRevLett.125.200505} {\bibfield
  {journal} {\bibinfo  {journal} {Phys. Rev. Lett.}\ }\textbf {\bibinfo
  {volume} {125}},\ \bibinfo {pages} {200505} (\bibinfo {year}
  {2020})}\BibitemShut {NoStop}%
\bibitem [{\citenamefont {Doherty}\ and\ \citenamefont
  {Jacobs}(1999)}]{Doherty1999}%
  \BibitemOpen
  \bibfield  {author} {\bibinfo {author} {\bibfnamefont {A.~C.}\ \bibnamefont
  {Doherty}}\ and\ \bibinfo {author} {\bibfnamefont {K.}~\bibnamefont
  {Jacobs}},\ }\href {\doibase 10.1103/PhysRevA.60.2700} {\bibfield  {journal}
  {\bibinfo  {journal} {Phys. Rev. A}\ }\textbf {\bibinfo {volume} {60}},\
  \bibinfo {pages} {2700} (\bibinfo {year} {1999})}\BibitemShut {NoStop}%
\bibitem [{\citenamefont {Wiseman}\ and\ \citenamefont
  {Milburn}(1993)}]{Wiseman1993}%
  \BibitemOpen
  \bibfield  {author} {\bibinfo {author} {\bibfnamefont {H.~M.}\ \bibnamefont
  {Wiseman}}\ and\ \bibinfo {author} {\bibfnamefont {G.~J.}\ \bibnamefont
  {Milburn}},\ }\href {\doibase 10.1103/PhysRevLett.70.548} {\bibfield
  {journal} {\bibinfo  {journal} {Phys. Rev. Lett.}\ }\textbf {\bibinfo
  {volume} {70}},\ \bibinfo {pages} {548} (\bibinfo {year} {1993})}\BibitemShut
  {NoStop}%
\bibitem [{\citenamefont {Wiseman}\ and\ \citenamefont
  {Milburn}(1994)}]{Wiseman1994}%
  \BibitemOpen
  \bibfield  {author} {\bibinfo {author} {\bibfnamefont {H.~M.}\ \bibnamefont
  {Wiseman}}\ and\ \bibinfo {author} {\bibfnamefont {G.~J.}\ \bibnamefont
  {Milburn}},\ }\href {\doibase 10.1103/PhysRevA.49.1350} {\bibfield  {journal}
  {\bibinfo  {journal} {Phys. Rev. A}\ }\textbf {\bibinfo {volume} {49}},\
  \bibinfo {pages} {1350} (\bibinfo {year} {1994})}\BibitemShut {NoStop}%
\bibitem [{\citenamefont {Thomsen}\ \emph {et~al.}(2002)\citenamefont
  {Thomsen}, \citenamefont {Mancini},\ and\ \citenamefont
  {Wiseman}}]{Thomsen2002}%
  \BibitemOpen
  \bibfield  {author} {\bibinfo {author} {\bibfnamefont {L.~K.}\ \bibnamefont
  {Thomsen}}, \bibinfo {author} {\bibfnamefont {S.}~\bibnamefont {Mancini}}, \
  and\ \bibinfo {author} {\bibfnamefont {H.~M.}\ \bibnamefont {Wiseman}},\
  }\href {\doibase 10.1103/PhysRevA.65.061801} {\bibfield  {journal} {\bibinfo
  {journal} {Phys. Rev. A}\ }\textbf {\bibinfo {volume} {65}},\ \bibinfo
  {pages} {061801} (\bibinfo {year} {2002})},\ \Eprint
  {http://arxiv.org/abs/quant-ph/0202028} {quant-ph/0202028} \BibitemShut
  {NoStop}%
\bibitem [{\citenamefont {Serafini}\ and\ \citenamefont
  {Mancini}(2010)}]{SerafozziMancini}%
  \BibitemOpen
  \bibfield  {author} {\bibinfo {author} {\bibfnamefont {A.}~\bibnamefont
  {Serafini}}\ and\ \bibinfo {author} {\bibfnamefont {S.}~\bibnamefont
  {Mancini}},\ }\href {\doibase 10.1103/PhysRevLett.104.220501} {\bibfield
  {journal} {\bibinfo  {journal} {Phys. Rev. Lett.}\ }\textbf {\bibinfo
  {volume} {104}},\ \bibinfo {pages} {220501} (\bibinfo {year}
  {2010})}\BibitemShut {NoStop}%
\bibitem [{\citenamefont {Szorkovszky}\ \emph {et~al.}(2011)\citenamefont
  {Szorkovszky}, \citenamefont {Doherty}, \citenamefont {Harris},\ and\
  \citenamefont {Bowen}}]{Szorkovszky2011}%
  \BibitemOpen
  \bibfield  {author} {\bibinfo {author} {\bibfnamefont {A.}~\bibnamefont
  {Szorkovszky}}, \bibinfo {author} {\bibfnamefont {A.~C.}\ \bibnamefont
  {Doherty}}, \bibinfo {author} {\bibfnamefont {G.~I.}\ \bibnamefont {Harris}},
  \ and\ \bibinfo {author} {\bibfnamefont {W.~P.}\ \bibnamefont {Bowen}},\
  }\href {\doibase 10.1103/PhysRevLett.107.213603} {\bibfield  {journal}
  {\bibinfo  {journal} {Phys. Rev. Lett.}\ }\textbf {\bibinfo {volume} {107}},\
  \bibinfo {pages} {213603} (\bibinfo {year} {2011})}\BibitemShut {NoStop}%
\bibitem [{\citenamefont {Genoni}\ \emph {et~al.}(2013)\citenamefont {Genoni},
  \citenamefont {Mancini},\ and\ \citenamefont {Serafini}}]{Genoni2013PRA}%
  \BibitemOpen
  \bibfield  {author} {\bibinfo {author} {\bibfnamefont {M.~G.}\ \bibnamefont
  {Genoni}}, \bibinfo {author} {\bibfnamefont {S.}~\bibnamefont {Mancini}}, \
  and\ \bibinfo {author} {\bibfnamefont {A.}~\bibnamefont {Serafini}},\ }\href
  {\doibase 10.1103/PhysRevA.87.042333} {\bibfield  {journal} {\bibinfo
  {journal} {Phys. Rev. A}\ }\textbf {\bibinfo {volume} {87}},\ \bibinfo
  {pages} {042333} (\bibinfo {year} {2013})}\BibitemShut {NoStop}%
\bibitem [{\citenamefont {Genoni}\ \emph {et~al.}(2015)\citenamefont {Genoni},
  \citenamefont {Zhang}, \citenamefont {Millen}, \citenamefont {Barker},\ and\
  \citenamefont {Serafini}}]{Genoni2015NJP}%
  \BibitemOpen
  \bibfield  {author} {\bibinfo {author} {\bibfnamefont {M.~G.}\ \bibnamefont
  {Genoni}}, \bibinfo {author} {\bibfnamefont {J.}~\bibnamefont {Zhang}},
  \bibinfo {author} {\bibfnamefont {J.}~\bibnamefont {Millen}}, \bibinfo
  {author} {\bibfnamefont {P.~F.}\ \bibnamefont {Barker}}, \ and\ \bibinfo
  {author} {\bibfnamefont {A.}~\bibnamefont {Serafini}},\ }\href {\doibase
  10.1088/1367-2630/17/7/073019} {\bibfield  {journal} {\bibinfo  {journal}
  {New Journal of Physics}\ }\textbf {\bibinfo {volume} {17}},\ \bibinfo
  {pages} {073019} (\bibinfo {year} {2015})}\BibitemShut {NoStop}%
\bibitem [{\citenamefont {Hofer}\ and\ \citenamefont
  {Hammerer}(2015)}]{Hofer2015}%
  \BibitemOpen
  \bibfield  {author} {\bibinfo {author} {\bibfnamefont {S.~G.}\ \bibnamefont
  {Hofer}}\ and\ \bibinfo {author} {\bibfnamefont {K.}~\bibnamefont
  {Hammerer}},\ }\href {\doibase 10.1103/PhysRevA.91.033822} {\bibfield
  {journal} {\bibinfo  {journal} {Phys. Rev. A}\ }\textbf {\bibinfo {volume}
  {91}},\ \bibinfo {pages} {033822} (\bibinfo {year} {2015})},\ \Eprint
  {http://arxiv.org/abs/1411.1337} {1411.1337} \BibitemShut {NoStop}%
\bibitem [{\citenamefont {Brunelli}\ \emph {et~al.}(2019)\citenamefont
  {Brunelli}, \citenamefont {Malz},\ and\ \citenamefont
  {Nunnenkamp}}]{Brunelli2019PRL}%
  \BibitemOpen
  \bibfield  {author} {\bibinfo {author} {\bibfnamefont {M.}~\bibnamefont
  {Brunelli}}, \bibinfo {author} {\bibfnamefont {D.}~\bibnamefont {Malz}}, \
  and\ \bibinfo {author} {\bibfnamefont {A.}~\bibnamefont {Nunnenkamp}},\
  }\href {\doibase 10.1103/PhysRevLett.123.093602} {\bibfield  {journal}
  {\bibinfo  {journal} {Phys. Rev. Lett.}\ }\textbf {\bibinfo {volume} {123}},\
  \bibinfo {pages} {093602} (\bibinfo {year} {2019})}\BibitemShut {NoStop}%
\bibitem [{\citenamefont {Di~Giovanni}\ \emph {et~al.}(2021)\citenamefont
  {Di~Giovanni}, \citenamefont {Brunelli},\ and\ \citenamefont
  {Genoni}}]{DiGiovanni2021}%
  \BibitemOpen
  \bibfield  {author} {\bibinfo {author} {\bibfnamefont {A.}~\bibnamefont
  {Di~Giovanni}}, \bibinfo {author} {\bibfnamefont {M.}~\bibnamefont
  {Brunelli}}, \ and\ \bibinfo {author} {\bibfnamefont {M.~G.}\ \bibnamefont
  {Genoni}},\ }\href {\doibase 10.1103/PhysRevA.103.022614} {\bibfield
  {journal} {\bibinfo  {journal} {Phys. Rev. A}\ }\textbf {\bibinfo {volume}
  {103}},\ \bibinfo {pages} {022614} (\bibinfo {year} {2021})}\BibitemShut
  {NoStop}%
\bibitem [{\citenamefont {Rossi}\ \emph {et~al.}(2018)\citenamefont {Rossi},
  \citenamefont {Mason}, \citenamefont {Chen}, \citenamefont {Tsaturyan},\ and\
  \citenamefont {Schliesser}}]{Rossi2018}%
  \BibitemOpen
  \bibfield  {author} {\bibinfo {author} {\bibfnamefont {M.}~\bibnamefont
  {Rossi}}, \bibinfo {author} {\bibfnamefont {D.}~\bibnamefont {Mason}},
  \bibinfo {author} {\bibfnamefont {J.}~\bibnamefont {Chen}}, \bibinfo {author}
  {\bibfnamefont {Y.}~\bibnamefont {Tsaturyan}}, \ and\ \bibinfo {author}
  {\bibfnamefont {A.}~\bibnamefont {Schliesser}},\ }\href {\doibase
  10.1038/s41586-018-0643-8} {\bibfield  {journal} {\bibinfo  {journal}
  {Nature}\ }\textbf {\bibinfo {volume} {563}},\ \bibinfo {pages} {53}
  (\bibinfo {year} {2018})}\BibitemShut {NoStop}%
\bibitem [{\citenamefont {Magrini}\ \emph {et~al.}(2021)\citenamefont
  {Magrini}, \citenamefont {Rosenzweig}, \citenamefont {Bach}, \citenamefont
  {Deutschmann-Olek}, \citenamefont {Hofer}, \citenamefont {Hong},
  \citenamefont {Kiesel}, \citenamefont {Kugi},\ and\ \citenamefont
  {Aspelmeyer}}]{Magrini2021}%
  \BibitemOpen
  \bibfield  {author} {\bibinfo {author} {\bibfnamefont {L.}~\bibnamefont
  {Magrini}}, \bibinfo {author} {\bibfnamefont {P.}~\bibnamefont {Rosenzweig}},
  \bibinfo {author} {\bibfnamefont {C.}~\bibnamefont {Bach}}, \bibinfo {author}
  {\bibfnamefont {A.}~\bibnamefont {Deutschmann-Olek}}, \bibinfo {author}
  {\bibfnamefont {S.~G.}\ \bibnamefont {Hofer}}, \bibinfo {author}
  {\bibfnamefont {S.}~\bibnamefont {Hong}}, \bibinfo {author} {\bibfnamefont
  {N.}~\bibnamefont {Kiesel}}, \bibinfo {author} {\bibfnamefont
  {A.}~\bibnamefont {Kugi}}, \ and\ \bibinfo {author} {\bibfnamefont
  {M.}~\bibnamefont {Aspelmeyer}},\ }\href {\doibase
  10.1038/s41586-021-03602-3} {\bibfield  {journal} {\bibinfo  {journal}
  {Nature}\ }\textbf {\bibinfo {volume} {595}},\ \bibinfo {pages} {373}
  (\bibinfo {year} {2021})}\BibitemShut {NoStop}%
\bibitem [{\citenamefont {Tebbenjohanns}\ \emph {et~al.}(2021)\citenamefont
  {Tebbenjohanns}, \citenamefont {Mattana}, \citenamefont {Rossi},
  \citenamefont {Frimmer},\ and\ \citenamefont {Novotny}}]{Tebbenjohanns2021}%
  \BibitemOpen
  \bibfield  {author} {\bibinfo {author} {\bibfnamefont {F.}~\bibnamefont
  {Tebbenjohanns}}, \bibinfo {author} {\bibfnamefont {M.~L.}\ \bibnamefont
  {Mattana}}, \bibinfo {author} {\bibfnamefont {M.}~\bibnamefont {Rossi}},
  \bibinfo {author} {\bibfnamefont {M.}~\bibnamefont {Frimmer}}, \ and\
  \bibinfo {author} {\bibfnamefont {L.}~\bibnamefont {Novotny}},\ }\href
  {\doibase 10.1038/s41586-021-03617-w} {\bibfield  {journal} {\bibinfo
  {journal} {Nature}\ }\textbf {\bibinfo {volume} {595}},\ \bibinfo {pages}
  {378} (\bibinfo {year} {2021})}\BibitemShut {NoStop}%
\bibitem [{\citenamefont {Sutton}\ and\ \citenamefont
  {Barto}(2018)}]{sutton2018reinforcement}%
  \BibitemOpen
  \bibfield  {author} {\bibinfo {author} {\bibfnamefont {R.~S.}\ \bibnamefont
  {Sutton}}\ and\ \bibinfo {author} {\bibfnamefont {A.~G.}\ \bibnamefont
  {Barto}},\ }\href
  {https://mitpress.mit.edu/books/reinforcement-learning-second-edition} {\emph
  {\bibinfo {title} {Reinforcement learning: An introduction}}}\ (\bibinfo
  {publisher} {MIT press},\ \bibinfo {year} {2018})\BibitemShut {NoStop}%
\bibitem [{\citenamefont {Mnih}\ \emph {et~al.}(2015)\citenamefont {Mnih},
  \citenamefont {Kavukcuoglu}, \citenamefont {Silver}, \citenamefont {Rusu},
  \citenamefont {Veness}, \citenamefont {Bellemare}, \citenamefont {Graves},
  \citenamefont {Riedmiller}, \citenamefont {Fidjeland}, \citenamefont
  {Ostrovski}, \citenamefont {Petersen}, \citenamefont {Beattie}, \citenamefont
  {Sadik}, \citenamefont {Antonoglou}, \citenamefont {King}, \citenamefont
  {Kumaran}, \citenamefont {Wierstra}, \citenamefont {Legg},\ and\
  \citenamefont {Hassabis}}]{deepmindgames2015}%
  \BibitemOpen
  \bibfield  {author} {\bibinfo {author} {\bibfnamefont {V.}~\bibnamefont
  {Mnih}}, \bibinfo {author} {\bibfnamefont {K.}~\bibnamefont {Kavukcuoglu}},
  \bibinfo {author} {\bibfnamefont {D.}~\bibnamefont {Silver}}, \bibinfo
  {author} {\bibfnamefont {A.~A.}\ \bibnamefont {Rusu}}, \bibinfo {author}
  {\bibfnamefont {J.}~\bibnamefont {Veness}}, \bibinfo {author} {\bibfnamefont
  {M.~G.}\ \bibnamefont {Bellemare}}, \bibinfo {author} {\bibfnamefont
  {A.}~\bibnamefont {Graves}}, \bibinfo {author} {\bibfnamefont
  {M.}~\bibnamefont {Riedmiller}}, \bibinfo {author} {\bibfnamefont {A.~K.}\
  \bibnamefont {Fidjeland}}, \bibinfo {author} {\bibfnamefont {G.}~\bibnamefont
  {Ostrovski}}, \bibinfo {author} {\bibfnamefont {S.}~\bibnamefont {Petersen}},
  \bibinfo {author} {\bibfnamefont {C.}~\bibnamefont {Beattie}}, \bibinfo
  {author} {\bibfnamefont {A.}~\bibnamefont {Sadik}}, \bibinfo {author}
  {\bibfnamefont {I.}~\bibnamefont {Antonoglou}}, \bibinfo {author}
  {\bibfnamefont {H.}~\bibnamefont {King}}, \bibinfo {author} {\bibfnamefont
  {D.}~\bibnamefont {Kumaran}}, \bibinfo {author} {\bibfnamefont
  {D.}~\bibnamefont {Wierstra}}, \bibinfo {author} {\bibfnamefont
  {S.}~\bibnamefont {Legg}}, \ and\ \bibinfo {author} {\bibfnamefont
  {D.}~\bibnamefont {Hassabis}},\ }\href {\doibase 10.1038/nature14236}
  {\bibfield  {journal} {\bibinfo  {journal} {Nature}\ }\textbf {\bibinfo
  {volume} {518}},\ \bibinfo {pages} {529} (\bibinfo {year}
  {2015})}\BibitemShut {NoStop}%
\bibitem [{\citenamefont {Silver}\ \emph {et~al.}(2017)\citenamefont {Silver},
  \citenamefont {Schrittwieser}, \citenamefont {Simonyan}, \citenamefont
  {Antonoglou}, \citenamefont {Huang}, \citenamefont {Guez}, \citenamefont
  {Hubert}, \citenamefont {Baker}, \citenamefont {Lai}, \citenamefont {Bolton},
  \citenamefont {Chen}, \citenamefont {Lillicrap}, \citenamefont {Hui},
  \citenamefont {Sifre}, \citenamefont {van~den Driessche}, \citenamefont
  {Graepel},\ and\ \citenamefont {Hassabis}}]{go2017nature}%
  \BibitemOpen
  \bibfield  {author} {\bibinfo {author} {\bibfnamefont {D.}~\bibnamefont
  {Silver}}, \bibinfo {author} {\bibfnamefont {J.}~\bibnamefont
  {Schrittwieser}}, \bibinfo {author} {\bibfnamefont {K.}~\bibnamefont
  {Simonyan}}, \bibinfo {author} {\bibfnamefont {I.}~\bibnamefont
  {Antonoglou}}, \bibinfo {author} {\bibfnamefont {A.}~\bibnamefont {Huang}},
  \bibinfo {author} {\bibfnamefont {A.}~\bibnamefont {Guez}}, \bibinfo {author}
  {\bibfnamefont {T.}~\bibnamefont {Hubert}}, \bibinfo {author} {\bibfnamefont
  {L.}~\bibnamefont {Baker}}, \bibinfo {author} {\bibfnamefont
  {M.}~\bibnamefont {Lai}}, \bibinfo {author} {\bibfnamefont {A.}~\bibnamefont
  {Bolton}}, \bibinfo {author} {\bibfnamefont {Y.}~\bibnamefont {Chen}},
  \bibinfo {author} {\bibfnamefont {T.}~\bibnamefont {Lillicrap}}, \bibinfo
  {author} {\bibfnamefont {F.}~\bibnamefont {Hui}}, \bibinfo {author}
  {\bibfnamefont {L.}~\bibnamefont {Sifre}}, \bibinfo {author} {\bibfnamefont
  {G.}~\bibnamefont {van~den Driessche}}, \bibinfo {author} {\bibfnamefont
  {T.}~\bibnamefont {Graepel}}, \ and\ \bibinfo {author} {\bibfnamefont
  {D.}~\bibnamefont {Hassabis}},\ }\href {\doibase 10.1038/nature24270}
  {\bibfield  {journal} {\bibinfo  {journal} {Nature}\ }\textbf {\bibinfo
  {volume} {550}},\ \bibinfo {pages} {354} (\bibinfo {year}
  {2017})}\BibitemShut {NoStop}%
\bibitem [{\citenamefont {Marquardt}(2021)}]{MarquardtTutorial}%
  \BibitemOpen
  \bibfield  {author} {\bibinfo {author} {\bibfnamefont {F.}~\bibnamefont
  {Marquardt}},\ }\href {\doibase 10.21468/SciPostPhysLectNotes.29} {\bibfield
  {journal} {\bibinfo  {journal} {SciPost Phys. Lect. Notes}\ ,\ \bibinfo
  {pages} {29}} (\bibinfo {year} {2021})}\BibitemShut {NoStop}%
\bibitem [{\citenamefont {F\"osel}\ \emph {et~al.}(2018)\citenamefont
  {F\"osel}, \citenamefont {Tighineanu}, \citenamefont {Weiss},\ and\
  \citenamefont {Marquardt}}]{Fosel2018}%
  \BibitemOpen
  \bibfield  {author} {\bibinfo {author} {\bibfnamefont {T.}~\bibnamefont
  {F\"osel}}, \bibinfo {author} {\bibfnamefont {P.}~\bibnamefont {Tighineanu}},
  \bibinfo {author} {\bibfnamefont {T.}~\bibnamefont {Weiss}}, \ and\ \bibinfo
  {author} {\bibfnamefont {F.}~\bibnamefont {Marquardt}},\ }\href {\doibase
  10.1103/PhysRevX.8.031084} {\bibfield  {journal} {\bibinfo  {journal} {Phys.
  Rev. X}\ }\textbf {\bibinfo {volume} {8}},\ \bibinfo {pages} {031084}
  (\bibinfo {year} {2018})}\BibitemShut {NoStop}%
\bibitem [{\citenamefont {Mavadia}\ \emph {et~al.}(2017)\citenamefont
  {Mavadia}, \citenamefont {Frey}, \citenamefont {Sastrawan}, \citenamefont
  {Dona},\ and\ \citenamefont {Biercuk}}]{Mavadia2017}%
  \BibitemOpen
  \bibfield  {author} {\bibinfo {author} {\bibfnamefont {S.}~\bibnamefont
  {Mavadia}}, \bibinfo {author} {\bibfnamefont {V.}~\bibnamefont {Frey}},
  \bibinfo {author} {\bibfnamefont {J.}~\bibnamefont {Sastrawan}}, \bibinfo
  {author} {\bibfnamefont {S.}~\bibnamefont {Dona}}, \ and\ \bibinfo {author}
  {\bibfnamefont {M.~J.}\ \bibnamefont {Biercuk}},\ }\href {\doibase
  10.1038/ncomms14106} {\bibfield  {journal} {\bibinfo  {journal} {Nature
  Communications}\ }\textbf {\bibinfo {volume} {8}},\ \bibinfo {pages} {14106}
  (\bibinfo {year} {2017})}\BibitemShut {NoStop}%
\bibitem [{\citenamefont {Niu}\ \emph {et~al.}(2019)\citenamefont {Niu},
  \citenamefont {Boixo}, \citenamefont {Smelyanskiy},\ and\ \citenamefont
  {Neven}}]{Niu2019}%
  \BibitemOpen
  \bibfield  {author} {\bibinfo {author} {\bibfnamefont {M.~Y.}\ \bibnamefont
  {Niu}}, \bibinfo {author} {\bibfnamefont {S.}~\bibnamefont {Boixo}}, \bibinfo
  {author} {\bibfnamefont {V.~N.}\ \bibnamefont {Smelyanskiy}}, \ and\ \bibinfo
  {author} {\bibfnamefont {H.}~\bibnamefont {Neven}},\ }\href {\doibase
  10.1038/s41534-019-0141-3} {\bibfield  {journal} {\bibinfo  {journal} {npj
  Quantum Information}\ }\textbf {\bibinfo {volume} {5}},\ \bibinfo {pages}
  {33} (\bibinfo {year} {2019})}\BibitemShut {NoStop}%
\bibitem [{\citenamefont {Porotti}\ \emph {et~al.}(2019)\citenamefont
  {Porotti}, \citenamefont {Tamascelli}, \citenamefont {Restelli},\ and\
  \citenamefont {Prati}}]{Porotti2019}%
  \BibitemOpen
  \bibfield  {author} {\bibinfo {author} {\bibfnamefont {R.}~\bibnamefont
  {Porotti}}, \bibinfo {author} {\bibfnamefont {D.}~\bibnamefont {Tamascelli}},
  \bibinfo {author} {\bibfnamefont {M.}~\bibnamefont {Restelli}}, \ and\
  \bibinfo {author} {\bibfnamefont {E.}~\bibnamefont {Prati}},\ }\href
  {\doibase 10.1038/s42005-019-0169-x} {\bibfield  {journal} {\bibinfo
  {journal} {Communications Physics}\ }\textbf {\bibinfo {volume} {2}},\
  \bibinfo {pages} {61} (\bibinfo {year} {2019})}\BibitemShut {NoStop}%
\bibitem [{\citenamefont {Brown}\ \emph {et~al.}(2021)\citenamefont {Brown},
  \citenamefont {Sgroi}, \citenamefont {Giannelli}, \citenamefont {Paraoanu},
  \citenamefont {Paladino}, \citenamefont {Falci}, \citenamefont
  {Paternostro},\ and\ \citenamefont {Ferraro}}]{Brown2021}%
  \BibitemOpen
  \bibfield  {author} {\bibinfo {author} {\bibfnamefont {J.}~\bibnamefont
  {Brown}}, \bibinfo {author} {\bibfnamefont {P.}~\bibnamefont {Sgroi}},
  \bibinfo {author} {\bibfnamefont {L.}~\bibnamefont {Giannelli}}, \bibinfo
  {author} {\bibfnamefont {G.~S.}\ \bibnamefont {Paraoanu}}, \bibinfo {author}
  {\bibfnamefont {E.}~\bibnamefont {Paladino}}, \bibinfo {author}
  {\bibfnamefont {G.}~\bibnamefont {Falci}}, \bibinfo {author} {\bibfnamefont
  {M.}~\bibnamefont {Paternostro}}, \ and\ \bibinfo {author} {\bibfnamefont
  {A.}~\bibnamefont {Ferraro}},\ }\href@noop {} {\enquote {\bibinfo {title}
  {Reinforcement learning-enhanced protocols for coherent population-transfer
  in three-level quantum systems},}\ } (\bibinfo {year} {2021}),\ \Eprint
  {http://arxiv.org/abs/2109.00973} {arXiv:2109.00973 [quant-ph]} \BibitemShut
  {NoStop}%
\bibitem [{\citenamefont {Moro}\ \emph {et~al.}(2021)\citenamefont {Moro},
  \citenamefont {Paris}, \citenamefont {Restelli},\ and\ \citenamefont
  {Prati}}]{Moro2021}%
  \BibitemOpen
  \bibfield  {author} {\bibinfo {author} {\bibfnamefont {L.}~\bibnamefont
  {Moro}}, \bibinfo {author} {\bibfnamefont {M.~G.~A.}\ \bibnamefont {Paris}},
  \bibinfo {author} {\bibfnamefont {M.}~\bibnamefont {Restelli}}, \ and\
  \bibinfo {author} {\bibfnamefont {E.}~\bibnamefont {Prati}},\ }\href
  {\doibase 10.1038/s42005-021-00684-3} {\bibfield  {journal} {\bibinfo
  {journal} {Communications Physics}\ }\textbf {\bibinfo {volume} {4}},\
  \bibinfo {pages} {178} (\bibinfo {year} {2021})}\BibitemShut {NoStop}%
\bibitem [{\citenamefont {Corli}\ \emph {et~al.}(2021)\citenamefont {Corli},
  \citenamefont {Moro}, \citenamefont {Galli},\ and\ \citenamefont
  {Prati}}]{Corli2021}%
  \BibitemOpen
  \bibfield  {author} {\bibinfo {author} {\bibfnamefont {S.}~\bibnamefont
  {Corli}}, \bibinfo {author} {\bibfnamefont {L.}~\bibnamefont {Moro}},
  \bibinfo {author} {\bibfnamefont {D.~E.}\ \bibnamefont {Galli}}, \ and\
  \bibinfo {author} {\bibfnamefont {E.}~\bibnamefont {Prati}},\ }\href
  {\doibase 10.1088/1751-8121/ac2596} {\bibfield  {journal} {\bibinfo
  {journal} {Journal of Physics A: Mathematical and Theoretical}\ }\textbf
  {\bibinfo {volume} {54}},\ \bibinfo {pages} {425302} (\bibinfo {year}
  {2021})}\BibitemShut {NoStop}%
\bibitem [{\citenamefont {Borah}\ \emph {et~al.}(2021)\citenamefont {Borah},
  \citenamefont {Sarma}, \citenamefont {Kewming}, \citenamefont {Milburn},\
  and\ \citenamefont {Twamley}}]{borah2021}%
  \BibitemOpen
  \bibfield  {author} {\bibinfo {author} {\bibfnamefont {S.}~\bibnamefont
  {Borah}}, \bibinfo {author} {\bibfnamefont {B.}~\bibnamefont {Sarma}},
  \bibinfo {author} {\bibfnamefont {M.}~\bibnamefont {Kewming}}, \bibinfo
  {author} {\bibfnamefont {G.~J.}\ \bibnamefont {Milburn}}, \ and\ \bibinfo
  {author} {\bibfnamefont {J.}~\bibnamefont {Twamley}},\ }\href
  {http://arxiv.org/abs/2104.11856} {\bibfield  {journal} {\bibinfo  {journal}
  {arXiv:2104.11856 [physics, physics:quant-ph]}\ } (\bibinfo {year} {2021})},\
  \bibinfo {note} {arXiv: 2104.11856}\BibitemShut {NoStop}%
\bibitem [{\citenamefont {Porotti}\ \emph {et~al.}(2021)\citenamefont
  {Porotti}, \citenamefont {Essig}, \citenamefont {Huard},\ and\ \citenamefont
  {Marquardt}}]{Porotti2021}%
  \BibitemOpen
  \bibfield  {author} {\bibinfo {author} {\bibfnamefont {R.}~\bibnamefont
  {Porotti}}, \bibinfo {author} {\bibfnamefont {A.}~\bibnamefont {Essig}},
  \bibinfo {author} {\bibfnamefont {B.}~\bibnamefont {Huard}}, \ and\ \bibinfo
  {author} {\bibfnamefont {F.}~\bibnamefont {Marquardt}},\ }\href@noop {}
  {\enquote {\bibinfo {title} {Deep reinforcement learning for quantum state
  preparation with weak nonlinear measurements},}\ } (\bibinfo {year} {2021}),\
  \Eprint {http://arxiv.org/abs/2107.08816} {arXiv:2107.08816 [quant-ph]}
  \BibitemShut {NoStop}%
\bibitem [{\citenamefont {Evans}\ \emph {et~al.}(2021)\citenamefont {Evans},
  \citenamefont {Wang}, \citenamefont {Frim}, \citenamefont {DeWeese},\ and\
  \citenamefont {Theodorou}}]{evans2021stochastic}%
  \BibitemOpen
  \bibfield  {author} {\bibinfo {author} {\bibfnamefont {E.~N.}\ \bibnamefont
  {Evans}}, \bibinfo {author} {\bibfnamefont {Z.}~\bibnamefont {Wang}},
  \bibinfo {author} {\bibfnamefont {A.~G.}\ \bibnamefont {Frim}}, \bibinfo
  {author} {\bibfnamefont {M.~R.}\ \bibnamefont {DeWeese}}, \ and\ \bibinfo
  {author} {\bibfnamefont {E.~A.}\ \bibnamefont {Theodorou}},\ }\href@noop {}
  {\enquote {\bibinfo {title} {Stochastic optimization for learning quantum
  state feedback control},}\ } (\bibinfo {year} {2021}),\ \Eprint
  {http://arxiv.org/abs/2111.09896} {arXiv:2111.09896 [quant-ph]} \BibitemShut
  {NoStop}%
\bibitem [{\citenamefont {Serafini}(2017)}]{SerafozziBook}%
  \BibitemOpen
  \bibfield  {author} {\bibinfo {author} {\bibfnamefont {A.}~\bibnamefont
  {Serafini}},\ }\href@noop {} {\emph {\bibinfo {title} {{Quantum Continuous
  Variables}}}}\ (\bibinfo  {publisher} {CRC Press},\ \bibinfo {address} {Boca
  Raton},\ \bibinfo {year} {2017})\BibitemShut {NoStop}%
\bibitem [{\citenamefont {Candia}\ \emph {et~al.}(2021)\citenamefont {Candia},
  \citenamefont {Minganti}, \citenamefont {Petrovnin}, \citenamefont
  {Paraoanu},\ and\ \citenamefont {Felicetti}}]{Dicandia2021}%
  \BibitemOpen
  \bibfield  {author} {\bibinfo {author} {\bibfnamefont {R.~D.}\ \bibnamefont
  {Candia}}, \bibinfo {author} {\bibfnamefont {F.}~\bibnamefont {Minganti}},
  \bibinfo {author} {\bibfnamefont {K.~V.}\ \bibnamefont {Petrovnin}}, \bibinfo
  {author} {\bibfnamefont {G.~S.}\ \bibnamefont {Paraoanu}}, \ and\ \bibinfo
  {author} {\bibfnamefont {S.}~\bibnamefont {Felicetti}},\ }\href@noop {}
  {\enquote {\bibinfo {title} {Critical parametric quantum sensing},}\ }
  (\bibinfo {year} {2021}),\ \Eprint {http://arxiv.org/abs/2107.04503}
  {arXiv:2107.04503 [quant-ph]} \BibitemShut {NoStop}%
\bibitem [{\citenamefont {Monras}(2006)}]{Monras2006}%
  \BibitemOpen
  \bibfield  {author} {\bibinfo {author} {\bibfnamefont {A.}~\bibnamefont
  {Monras}},\ }\href {\doibase 10.1103/PhysRevA.73.033821} {\bibfield
  {journal} {\bibinfo  {journal} {Phys. Rev. A}\ }\textbf {\bibinfo {volume}
  {73}},\ \bibinfo {pages} {033821} (\bibinfo {year} {2006})}\BibitemShut
  {NoStop}%
\bibitem [{\citenamefont {Genoni}\ \emph {et~al.}(2011)\citenamefont {Genoni},
  \citenamefont {Olivares},\ and\ \citenamefont {Paris}}]{GenoniPRL2011}%
  \BibitemOpen
  \bibfield  {author} {\bibinfo {author} {\bibfnamefont {M.~G.}\ \bibnamefont
  {Genoni}}, \bibinfo {author} {\bibfnamefont {S.}~\bibnamefont {Olivares}}, \
  and\ \bibinfo {author} {\bibfnamefont {M.~G.~A.}\ \bibnamefont {Paris}},\
  }\href {\doibase 10.1103/PhysRevLett.106.153603} {\bibfield  {journal}
  {\bibinfo  {journal} {Phys. Rev. Lett.}\ }\textbf {\bibinfo {volume} {106}},\
  \bibinfo {pages} {153603} (\bibinfo {year} {2011})}\BibitemShut {NoStop}%
\bibitem [{\citenamefont {Schleier-Smith}\ \emph {et~al.}(2010)\citenamefont
  {Schleier-Smith}, \citenamefont {Leroux},\ and\ \citenamefont
  {Vuleti\ifmmode~\acute{c}\else \'{c}\fi{}}}]{SchleierSmith2010}%
  \BibitemOpen
  \bibfield  {author} {\bibinfo {author} {\bibfnamefont {M.~H.}\ \bibnamefont
  {Schleier-Smith}}, \bibinfo {author} {\bibfnamefont {I.~D.}\ \bibnamefont
  {Leroux}}, \ and\ \bibinfo {author} {\bibfnamefont {V.}~\bibnamefont
  {Vuleti\ifmmode~\acute{c}\else \'{c}\fi{}}},\ }\href {\doibase
  10.1103/PhysRevLett.104.073604} {\bibfield  {journal} {\bibinfo  {journal}
  {Phys. Rev. Lett.}\ }\textbf {\bibinfo {volume} {104}},\ \bibinfo {pages}
  {073604} (\bibinfo {year} {2010})}\BibitemShut {NoStop}%
\bibitem [{\citenamefont {Leroux}\ \emph {et~al.}(2010)\citenamefont {Leroux},
  \citenamefont {Schleier-Smith},\ and\ \citenamefont
  {Vuleti\ifmmode~\acute{c}\else \'{c}\fi{}}}]{Leroux2010}%
  \BibitemOpen
  \bibfield  {author} {\bibinfo {author} {\bibfnamefont {I.~D.}\ \bibnamefont
  {Leroux}}, \bibinfo {author} {\bibfnamefont {M.~H.}\ \bibnamefont
  {Schleier-Smith}}, \ and\ \bibinfo {author} {\bibfnamefont {V.}~\bibnamefont
  {Vuleti\ifmmode~\acute{c}\else \'{c}\fi{}}},\ }\href {\doibase
  10.1103/PhysRevLett.104.073602} {\bibfield  {journal} {\bibinfo  {journal}
  {Phys. Rev. Lett.}\ }\textbf {\bibinfo {volume} {104}},\ \bibinfo {pages}
  {073602} (\bibinfo {year} {2010})}\BibitemShut {NoStop}%
\bibitem [{\citenamefont {Wasilewski}\ \emph {et~al.}(2010)\citenamefont
  {Wasilewski}, \citenamefont {Jensen}, \citenamefont {Krauter}, \citenamefont
  {Renema}, \citenamefont {Balabas},\ and\ \citenamefont
  {Polzik}}]{Wasilewski2010}%
  \BibitemOpen
  \bibfield  {author} {\bibinfo {author} {\bibfnamefont {W.}~\bibnamefont
  {Wasilewski}}, \bibinfo {author} {\bibfnamefont {K.}~\bibnamefont {Jensen}},
  \bibinfo {author} {\bibfnamefont {H.}~\bibnamefont {Krauter}}, \bibinfo
  {author} {\bibfnamefont {J.~J.}\ \bibnamefont {Renema}}, \bibinfo {author}
  {\bibfnamefont {M.~V.}\ \bibnamefont {Balabas}}, \ and\ \bibinfo {author}
  {\bibfnamefont {E.~S.}\ \bibnamefont {Polzik}},\ }\href {\doibase
  10.1103/PhysRevLett.104.133601} {\bibfield  {journal} {\bibinfo  {journal}
  {Phys. Rev. Lett.}\ }\textbf {\bibinfo {volume} {104}},\ \bibinfo {pages}
  {133601} (\bibinfo {year} {2010})}\BibitemShut {NoStop}%
\bibitem [{\citenamefont {Wiseman}\ and\ \citenamefont
  {Doherty}(2005)}]{WisemanDoherty}%
  \BibitemOpen
  \bibfield  {author} {\bibinfo {author} {\bibfnamefont {H.~M.}\ \bibnamefont
  {Wiseman}}\ and\ \bibinfo {author} {\bibfnamefont {A.~C.}\ \bibnamefont
  {Doherty}},\ }\href {\doibase 10.1103/PhysRevLett.94.070405} {\bibfield
  {journal} {\bibinfo  {journal} {Phys. Rev. Lett.}\ }\textbf {\bibinfo
  {volume} {94}},\ \bibinfo {pages} {070405} (\bibinfo {year} {2005})},\
  \Eprint {http://arxiv.org/abs/0408099v4} {0408099v4 [quant-ph]} \BibitemShut
  {NoStop}%
\bibitem [{\citenamefont {Genoni}\ \emph {et~al.}(2016)\citenamefont {Genoni},
  \citenamefont {Lami},\ and\ \citenamefont {Serafini}}]{Diffusione}%
  \BibitemOpen
  \bibfield  {author} {\bibinfo {author} {\bibfnamefont {M.~G.}\ \bibnamefont
  {Genoni}}, \bibinfo {author} {\bibfnamefont {L.}~\bibnamefont {Lami}}, \ and\
  \bibinfo {author} {\bibfnamefont {A.}~\bibnamefont {Serafini}},\ }\href
  {\doibase 10.1080/00107514.2015.1125624} {\bibfield  {journal} {\bibinfo
  {journal} {Contemp. Phys.}\ }\textbf {\bibinfo {volume} {57}},\ \bibinfo
  {pages} {331} (\bibinfo {year} {2016})}\BibitemShut {NoStop}%
\bibitem [{\citenamefont {Milburn}\ and\ \citenamefont
  {Walls}(1981)}]{MilburnWallsSqueezing}%
  \BibitemOpen
  \bibfield  {author} {\bibinfo {author} {\bibfnamefont {G.}~\bibnamefont
  {Milburn}}\ and\ \bibinfo {author} {\bibfnamefont {D.}~\bibnamefont
  {Walls}},\ }\href {\doibase https://doi.org/10.1016/0030-4018(81)90232-7}
  {\bibfield  {journal} {\bibinfo  {journal} {Optics Communications}\ }\textbf
  {\bibinfo {volume} {39}},\ \bibinfo {pages} {401} (\bibinfo {year}
  {1981})}\BibitemShut {NoStop}%
\bibitem [{\citenamefont {Collett}\ and\ \citenamefont
  {Gardiner}(1984)}]{CollettGardinerSqueezing}%
  \BibitemOpen
  \bibfield  {author} {\bibinfo {author} {\bibfnamefont {M.~J.}\ \bibnamefont
  {Collett}}\ and\ \bibinfo {author} {\bibfnamefont {C.~W.}\ \bibnamefont
  {Gardiner}},\ }\href {\doibase 10.1103/PhysRevA.30.1386} {\bibfield
  {journal} {\bibinfo  {journal} {Phys. Rev. A}\ }\textbf {\bibinfo {volume}
  {30}},\ \bibinfo {pages} {1386} (\bibinfo {year} {1984})}\BibitemShut
  {NoStop}%
\bibitem [{\citenamefont {Huelga}\ \emph {et~al.}(1997)\citenamefont {Huelga},
  \citenamefont {Macchiavello}, \citenamefont {Pellizzari}, \citenamefont
  {Ekert}, \citenamefont {Plenio},\ and\ \citenamefont {Cirac}}]{Huelga97}%
  \BibitemOpen
  \bibfield  {author} {\bibinfo {author} {\bibfnamefont {S.~F.}\ \bibnamefont
  {Huelga}}, \bibinfo {author} {\bibfnamefont {C.}~\bibnamefont
  {Macchiavello}}, \bibinfo {author} {\bibfnamefont {T.}~\bibnamefont
  {Pellizzari}}, \bibinfo {author} {\bibfnamefont {A.~K.}\ \bibnamefont
  {Ekert}}, \bibinfo {author} {\bibfnamefont {M.~B.}\ \bibnamefont {Plenio}}, \
  and\ \bibinfo {author} {\bibfnamefont {J.~I.}\ \bibnamefont {Cirac}},\ }\href
  {\doibase 10.1103/PhysRevLett.79.3865} {\bibfield  {journal} {\bibinfo
  {journal} {Phys. Rev. Lett.}\ }\textbf {\bibinfo {volume} {79}},\ \bibinfo
  {pages} {3865} (\bibinfo {year} {1997})},\ \Eprint
  {http://arxiv.org/abs/quant-ph/9707014} {quant-ph/9707014} \BibitemShut
  {NoStop}%
\bibitem [{\citenamefont {Gammelmark}\ and\ \citenamefont
  {M{\o}lmer}(2013{\natexlab{b}})}]{Gammelmark2013a}%
  \BibitemOpen
  \bibfield  {author} {\bibinfo {author} {\bibfnamefont {S.}~\bibnamefont
  {Gammelmark}}\ and\ \bibinfo {author} {\bibfnamefont {K.}~\bibnamefont
  {M{\o}lmer}},\ }\href {\doibase 10.1103/PhysRevA.87.032115} {\bibfield
  {journal} {\bibinfo  {journal} {Phys. Rev. A}\ }\textbf {\bibinfo {volume}
  {87}},\ \bibinfo {pages} {032115} (\bibinfo {year} {2013}{\natexlab{b}})},\
  \Eprint {http://arxiv.org/abs/1212.5700} {1212.5700} \BibitemShut {NoStop}%
\bibitem [{\citenamefont {Oh}\ \emph {et~al.}(2019)\citenamefont {Oh},
  \citenamefont {Lee}, \citenamefont {Rockstuhl}, \citenamefont {Jeong},
  \citenamefont {Kim}, \citenamefont {Nha},\ and\ \citenamefont
  {Lee}}]{Oh2019}%
  \BibitemOpen
  \bibfield  {author} {\bibinfo {author} {\bibfnamefont {C.}~\bibnamefont
  {Oh}}, \bibinfo {author} {\bibfnamefont {C.}~\bibnamefont {Lee}}, \bibinfo
  {author} {\bibfnamefont {C.}~\bibnamefont {Rockstuhl}}, \bibinfo {author}
  {\bibfnamefont {H.}~\bibnamefont {Jeong}}, \bibinfo {author} {\bibfnamefont
  {J.}~\bibnamefont {Kim}}, \bibinfo {author} {\bibfnamefont {H.}~\bibnamefont
  {Nha}}, \ and\ \bibinfo {author} {\bibfnamefont {S.-Y.}\ \bibnamefont
  {Lee}},\ }\href {\doibase 10.1038/s41534-019-0124-4} {\bibfield  {journal}
  {\bibinfo  {journal} {npj Quantum Information}\ }\textbf {\bibinfo {volume}
  {5}},\ \bibinfo {pages} {10} (\bibinfo {year} {2019})}\BibitemShut {NoStop}%
\bibitem [{\citenamefont {Paris}(2009)}]{MatteoIJQI}%
  \BibitemOpen
  \bibfield  {author} {\bibinfo {author} {\bibfnamefont {M.~G.~A.}\
  \bibnamefont {Paris}},\ }\href {\doibase 10.1142/S0219749909004839}
  {\bibfield  {journal} {\bibinfo  {journal} {Int. J. Quant. Inf.}\ }\textbf
  {\bibinfo {volume} {07}},\ \bibinfo {pages} {125} (\bibinfo {year}
  {2009})}\BibitemShut {NoStop}%
\bibitem [{\citenamefont {Schulman}\ \emph {et~al.}(2017)\citenamefont
  {Schulman}, \citenamefont {Wolski}, \citenamefont {Dhariwal}, \citenamefont
  {Radford},\ and\ \citenamefont {Klimov}}]{schulman2017proximal}%
  \BibitemOpen
  \bibfield  {author} {\bibinfo {author} {\bibfnamefont {J.}~\bibnamefont
  {Schulman}}, \bibinfo {author} {\bibfnamefont {F.}~\bibnamefont {Wolski}},
  \bibinfo {author} {\bibfnamefont {P.}~\bibnamefont {Dhariwal}}, \bibinfo
  {author} {\bibfnamefont {A.}~\bibnamefont {Radford}}, \ and\ \bibinfo
  {author} {\bibfnamefont {O.}~\bibnamefont {Klimov}},\ }\href
  {https://arxiv.org/abs/1707.06347} {\bibfield  {journal} {\bibinfo  {journal}
  {arXiv preprint arXiv:1707.06347}\ } (\bibinfo {year} {2017})}\BibitemShut
  {NoStop}%
\bibitem [{\citenamefont {Hill}\ \emph {et~al.}(2018)\citenamefont {Hill},
  \citenamefont {Raffin}, \citenamefont {Ernestus}, \citenamefont {Gleave},
  \citenamefont {Kanervisto}, \citenamefont {Traore}, \citenamefont {Dhariwal},
  \citenamefont {Hesse}, \citenamefont {Klimov}, \citenamefont {Nichol},
  \citenamefont {Plappert}, \citenamefont {Radford}, \citenamefont {Schulman},
  \citenamefont {Sidor},\ and\ \citenamefont {Wu}}]{stable-baselines}%
  \BibitemOpen
  \bibfield  {author} {\bibinfo {author} {\bibfnamefont {A.}~\bibnamefont
  {Hill}}, \bibinfo {author} {\bibfnamefont {A.}~\bibnamefont {Raffin}},
  \bibinfo {author} {\bibfnamefont {M.}~\bibnamefont {Ernestus}}, \bibinfo
  {author} {\bibfnamefont {A.}~\bibnamefont {Gleave}}, \bibinfo {author}
  {\bibfnamefont {A.}~\bibnamefont {Kanervisto}}, \bibinfo {author}
  {\bibfnamefont {R.}~\bibnamefont {Traore}}, \bibinfo {author} {\bibfnamefont
  {P.}~\bibnamefont {Dhariwal}}, \bibinfo {author} {\bibfnamefont
  {C.}~\bibnamefont {Hesse}}, \bibinfo {author} {\bibfnamefont
  {O.}~\bibnamefont {Klimov}}, \bibinfo {author} {\bibfnamefont
  {A.}~\bibnamefont {Nichol}}, \bibinfo {author} {\bibfnamefont
  {M.}~\bibnamefont {Plappert}}, \bibinfo {author} {\bibfnamefont
  {A.}~\bibnamefont {Radford}}, \bibinfo {author} {\bibfnamefont
  {J.}~\bibnamefont {Schulman}}, \bibinfo {author} {\bibfnamefont
  {S.}~\bibnamefont {Sidor}}, \ and\ \bibinfo {author} {\bibfnamefont
  {Y.}~\bibnamefont {Wu}},\ }\href@noop {} {\enquote {\bibinfo {title} {Stable
  baselines},}\ }\bibinfo {howpublished}
  {\url{https://github.com/hill-a/stable-baselines}} (\bibinfo {year}
  {2018})\BibitemShut {NoStop}%
\bibitem [{\citenamefont {Pinel}\ \emph {et~al.}(2013)\citenamefont {Pinel},
  \citenamefont {Jian}, \citenamefont {Treps}, \citenamefont {Fabre},\ and\
  \citenamefont {Braun}}]{Pinel2013}%
  \BibitemOpen
  \bibfield  {author} {\bibinfo {author} {\bibfnamefont {O.}~\bibnamefont
  {Pinel}}, \bibinfo {author} {\bibfnamefont {P.}~\bibnamefont {Jian}},
  \bibinfo {author} {\bibfnamefont {N.}~\bibnamefont {Treps}}, \bibinfo
  {author} {\bibfnamefont {C.}~\bibnamefont {Fabre}}, \ and\ \bibinfo {author}
  {\bibfnamefont {D.}~\bibnamefont {Braun}},\ }\href {\doibase
  10.1103/PhysRevA.88.040102} {\bibfield  {journal} {\bibinfo  {journal} {Phys.
  Rev. A}\ }\textbf {\bibinfo {volume} {88}},\ \bibinfo {pages} {040102}
  (\bibinfo {year} {2013})},\ \Eprint {http://arxiv.org/abs/arXiv:1307.4637v1}
  {arXiv:1307.4637v1} \BibitemShut {NoStop}%
\end{thebibliography}%
\appendix
\section{Quantum metrology with continuously monitored quantum system}
\label{a:QMetrCM}
We start by giving a basic introduction on quantum estimation theory. Let us consider a quantum statistical model, that is a family of quantum states $\varrho_{\omega}$ parametrized by a parameter $\omega$ that we want to estimate. We now suppose to repeat $M$ times a measurement, corresponding to a certain POVM $\{\Pi_x\}$, on the quantum state, and thus collecting a set of measurement outcomes $\{x_j\}_{j=1}^M$. One can prove that the precision of any unbiased estimator $\tilde{\omega}$, that is a map from the measurement outcomes $\{x_j\}$ to the range of parameters taken by the $\omega$ is lower bounded according the Cram\'er-Rao bound 
\begin{align}
    \delta \omega \geq \frac{1}{\sqrt{M \, \mathcal{F}[p(x|\omega)]}} \,,
\end{align}
where we have introduced the classical Fisher information
\begin{align}
    \mathcal{F}[p(x|\omega)] &= \sum_x \frac{\left(\partial_\omega p(x|\omega)\right)^2}{p(x|\omega)} \,,\\
    &= \mathbbm{E}_{p(x|\omega)} \left[ \left( \frac{\partial_\omega p(x|\omega)}{p(x|\omega)}\right)^2 \right] \,,
\end{align}
 and we have denoted with $p(x|\omega)=\Tr[\varrho_{\omega}\Pi_x]$ the probability of obtaining the outcome $x$ from the measurement.
One can further optimize over all the possible measurements (POVM) $\{\Pi_x\}$ that one can perform on the quantum state $\varrho_{\omega}$, obtaining the quantum Cram\'er-Rao bound
\begin{align}
       \delta \omega \geq \frac{1}{\sqrt{M \, \mathcal{F}[p(x|\omega)]}} \geq \frac{1}{\sqrt{M \,\mathcal{Q}[\varrho_\omega]}}\,,
\end{align}
where we have introduced the quantum Fisher information (QFI)
\begin{align}
    \mathcal{Q}[\varrho_\omega] &= \Tr[\varrho_\omega L_\omega^2] \,,
\end{align}
written in terms of the symmetric logarithmic derivative defined via the Lyapunov equation
\begin{align}
    \frac{\partial\varrho_\omega}{\partial \omega} &= \frac{L_\omega \varrho_\omega + \varrho_\omega L_\omega}{2} \,.
\end{align}
Several alternative formulas can be derived for the QFI can be derived, based on the diagonalization of the state $\varrho_{\omega}$ or based on the fidelity between states characterized by parameters $\omega$ differing by an infinitesimal value \cite{MatteoIJQI}.

\par
If we want to estimate a parameter in continuously monitored quantum system, at each run of the experiment we obtain a a continuous measurement output (for example in the case of continuous homodyne detection) $\tilde{y}_t$ with a certain probability distribution $p_{\sf hom} = p(\tilde{y}_t |\omega)$ and corresponding to a particular trajectory for the quantum conditional state of the system $\varrho_c$. In this framework one proves that the bound on the estimation precision can be written as \cite{Albarelli2017a}
\begin{align}
    \delta \omega \geq \frac{1}{\sqrt{M \left( \mathcal{F}_{\sf hom} + \mathbbm{E}_{\sf traj}\left[\mathcal{Q}[\varrho_c] \right] \right)}} \,.
\end{align}
The relevant figure of merit it thus the effective QFI
\begin{align} 
\mathcal{Q}_{\sf eff} = \mathcal{F}_{\sf hom}  + \mathbbm{E}_{\sf traj}\left[\mathcal{Q}[\varrho_c] \right] \,,
\label{eq:effQFIapp}
\end{align}
corresponding to the sum of the Fisher information quantifying the information obtainable from the continuous homodyne results, plus the average of the quantum Fisher information of the conditional states, quantifying the information obtainable from a final measurement on $\varrho_c$. 

\section{Gaussian conditional dynamics and numerical evaluation of the effective QFI}  
\label{a:GaussianMonitored}
We here briefly review how to treat the evolution of continuously monitored quantum Gaussian states by following the approach in \cite{SerafozziBook,Diffusione} and how to evaluate the different figures of merit relevant for our purposes.

A Gaussian quantum state $\varrho$ of a continuous-variable quantum system is completely identified by its first moment vector $\bar{\bf r} = \Tr[\varrho\hat{\bf r}]$, and its covariance matrix $\boldsymbol{\sigma} = \Tr[\varrho_c \{\hat{\bf r} - \bar{\bf r}, (\hat{\bf r} - \bar{\bf r})^{\sf T} \} ]$, where $\{ \hat{\bf a}, \hat{\bf b}\} = \hat{\bf a} \hat{\bf b}^{\sf T} + (\hat{\bf b} \hat{\bf a}^{\sf T})^{\sf T}$. We remind that with this definition the covariance matrix for a single-mode system reads
\begin{align}
    \boldsymbol{\sigma} = 2 \left(
    \begin{array}{c c}
    \langle \Delta \hat{q}^2 \rangle & \langle \Delta \hat{q} \hat{p} \rangle \\
    \langle \Delta \hat{q} \hat{p} \rangle & \langle \Delta \hat{p}^2 \rangle
    \end{array}
    \right)
\end{align}
where 
\begin{align}
\langle \Delta \hat{A} \hat{B} \rangle =\Tr\left[\varrho\left(\frac{\hat{A}\hat{B} + \hat{B}\hat{A}}{2}\right)\right] - \Tr[\varrho\hat{A}]\Tr[\varrho\hat{B}]  
\end{align}
The covariance matrix thus directly contains all the squeezing properties of the quantum state $\varrho$.

\par
As we mentioned in the main text, the dynamics induced by the stochastic master equation (\ref{eq:sme}) preserves the Gaussian character of the quantum state, and the corresponding evolution is described by the equations
\begin{align}
d\bar{\bf r}_c &= A \bar{\bf r}_c \,dt + (E - \boldsymbol{\sigma}_c B) \frac{{\bf dw}_t}{\sqrt{2}} \,, 
\label{eq:rcApp}
\\
\frac{d\boldsymbol{\sigma}_c}{dt} &= A \boldsymbol{\sigma}_c + \boldsymbol{\sigma}_c A^{\sf T} + D - (E - \boldsymbol{\sigma}_c B)(E - \boldsymbol{\sigma}_c B)^{\sf T}, 
\label{eq:sigmacApp}
\end{align}
while the continuous homodyne outcome is written in vectorial form as
\begin{align} \label{eq:tamaB5}
{\bf dy}_t = -\sqrt{2} B^{\sf T} \bar{\bf r}_c \,dt + {\bf dw}_t\,.
\end{align}
\begin{figure*}
    \centering
    \includegraphics[width=.95\textwidth]{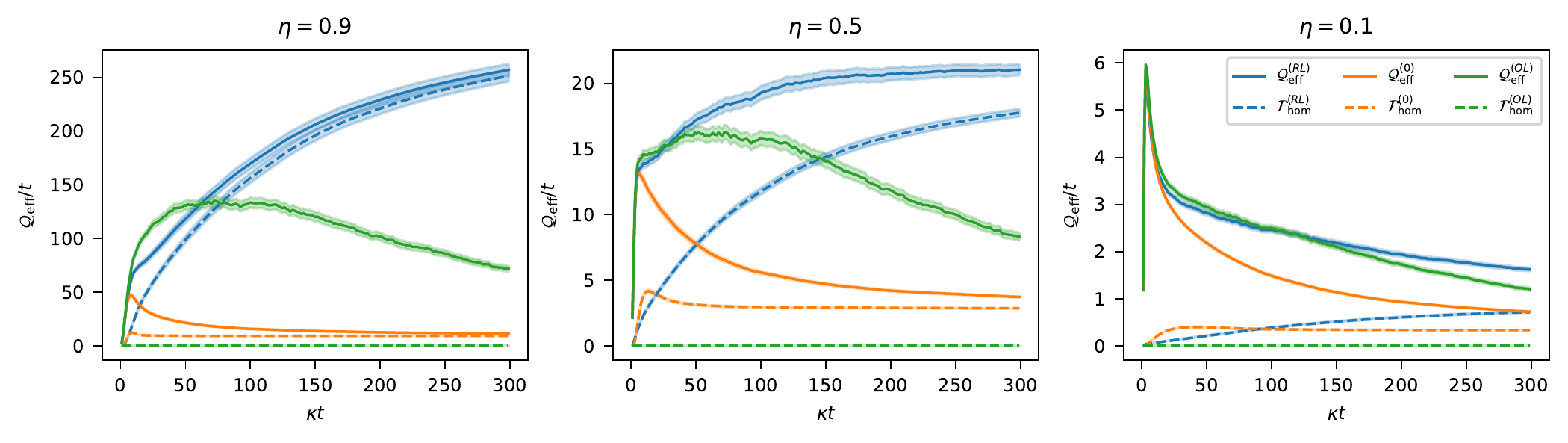}
    \caption{Performance of the control strategies as a function of time, quantified by the different Fisher informations divided by time, and for different values of $\eta$ (from left to right: $\eta = \{0.9,0.5,0.1\}$). All the other parameters are fixed as in Fig. \ref{f:compareQeff}.
    \label{f:compareQeffEta}}
\end{figure*}

The matrices entering in these equations can be derived by following different approaches~\cite{Diffusione,WisemanDoherty}, and for the physical setup we are interested in reads
\begin{align}
A&= 
\left(
\begin{array}{c c}
- (\chi + \kappa/2) & \omega \\
-\omega & \chi - \kappa/2 
\end{array}
\right)\,, \\
D &= \kappa \mathbbm{1}_2 \,, \\
B &= E =  \left(
\begin{array}{c c}
-\sqrt{\eta \kappa} & 0 \\
0 & 0 
\end{array}
\right) \,.
\end{align}
We observe that the matrices $B$ and $E$ are singular. The second component of the Wiener  increment ${\bf dw}_t$ in \eref{eq:tamaB5}, therefore, does not play any role at all, whereas the first component is determined by the homodyne detection output. 

The solution of the Riccati equation for the covariance matrix (\ref{eq:sigmac}) can be in general obtained numerically. However an analytical solution can be obtained for the steady-state covariance matrix for $\omega=0$ and by assuming a stable dynamics (that is $\chi<\kappa/2$), leading to
\begin{align}
\boldsymbol{\sigma}_c^{\sf ss}(\eta) = \left(
\begin{array}{c c}
\frac{\kappa(2 \eta-1) - 2 \chi + \sqrt{\kappa^2 - 4 \kappa\chi (2\eta-1) + 4 \chi^2}}{2\eta\kappa} & 0 \\
0 & \frac{\kappa}{\kappa - 2 \chi} 
\end{array}
\right) \,.
\label{eq:solutionsigmac}
\end{align}
Two opposite regimes can be observed here. By taking the limit for the efficiency $\eta \to 0$, and assuming a stable dynamics, we indeed obtain the solution for the unconditional (unmonitored) dynamics
\begin{align}
\boldsymbol{\sigma}_{\sf unc}^{\sf ss} = \left(
\begin{array}{c c}
\frac{\kappa}{\kappa + 2 \chi} & 0 \\
0 & \frac{\kappa}{\kappa - 2 \chi} 
\end{array}
\right) \,.
\label{eq:sigmacunc}
\end{align}

We thus find that for $0<\chi<\kappa/2$, the Hamiltonian squeezes the $\hat{q}$ quadrature, with a maximum amount of 3dB of squeezing near instability, that is for $\chi \to \kappa/2$~\cite{MilburnWallsSqueezing,CollettGardinerSqueezing}. In the opposite case of perfect monitoring, that is for $\eta=1$, we find
\begin{align}
\boldsymbol{\sigma}_c^{\sf ss} = \left(
\begin{array}{c c}
\frac{\kappa- 2 \chi }{\kappa} & 0 \\
0 & \frac{\kappa}{\kappa - 2 \chi} 
\end{array}
\right) \,,
\end{align}
which in turn, for $0<\chi<\kappa/2$, corresponds to even smaller variances of the $\hat{q}$ quadrature, and in principle infinite squeezing near instability. We remark that for $\omega\neq 0$ we numerically find that a lower amount of squeezing can be generated and that the most squeezed quadrature depends on the value of $\omega$ itself.

\begin{figure*}
    \includegraphics[width=\textwidth]{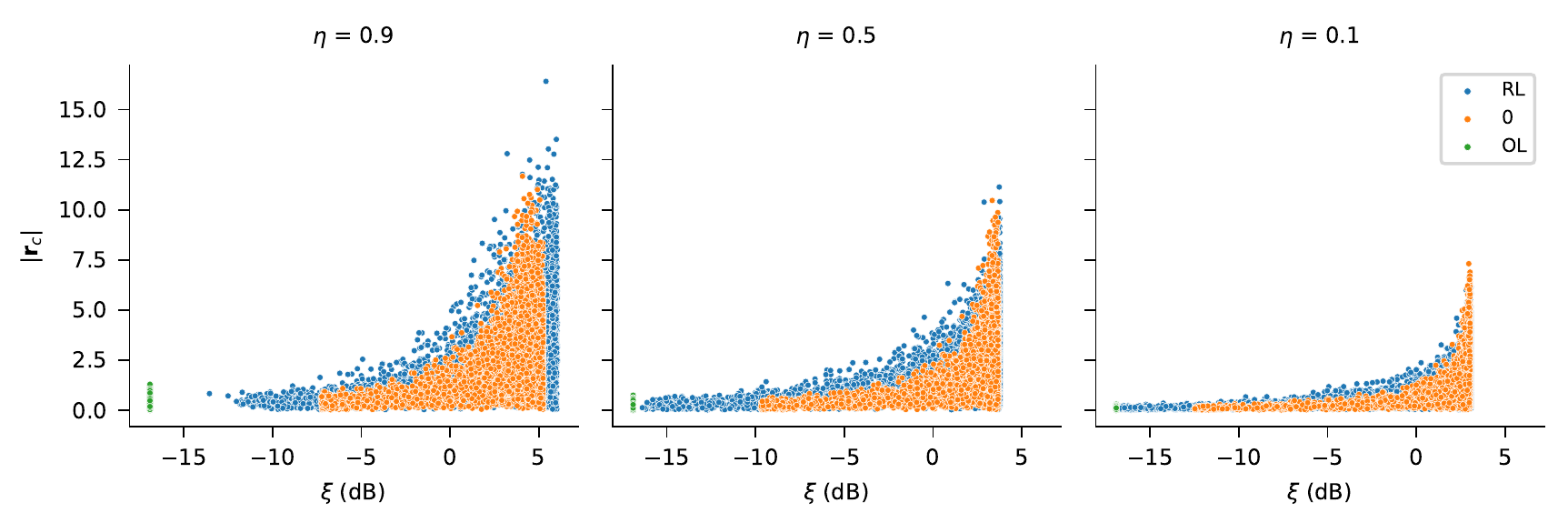}
    \caption{
    Scatter plot of the squeezing (expressed in dB) perpendicular to the conditional first moment vector $\bar{\mathbf{r}}_c$ and the absolute value of the first moment vector $|\bar{\mathbf{r}}_c|$, for the three control strategies at time $\kappa t = 180$ and for different values of $\eta$ (from left to right, $\eta = \{0.9, 0.5, 0.1\}$).}
    \label{f:scatterSqFM}
\end{figure*}

\begin{figure*}
    \includegraphics[width=\textwidth]{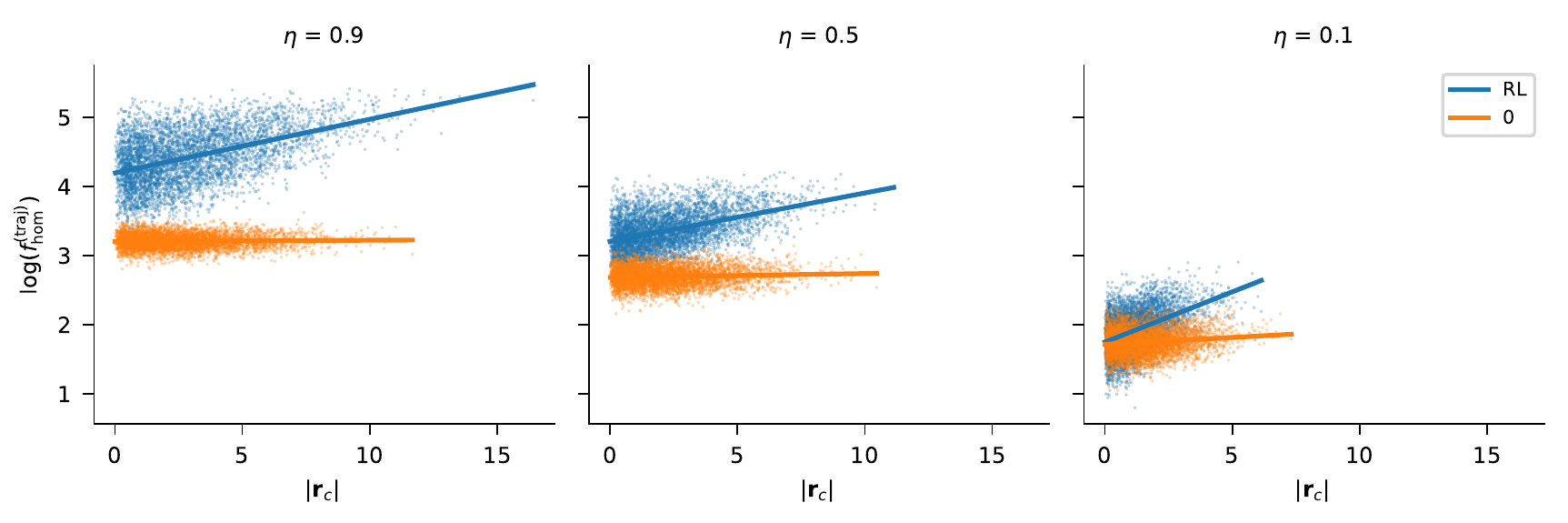}
    \caption{
    Scatter plot between the absolute value of the first moment vector $|\bar{\mathbf{r}}_c|$ and the trajectory-contribution to the homodyne Fisher information $\log(f_{\sf hom}^{\sf (traj)})$, at time $\kappa t = 180$ and for different values of $\eta$ (from left to right, $\eta = \{0.9, 0.5, 0.1\}$). We plot only the points corresponding to the agent and the no-control strategy, as $f_{\sf hom}^{\sf (traj)}=0$ for the OL-strategy.}
    \label{f:scatterFMfHom}
\end{figure*}

As we described in the previous section, the perfomance of the metrological protocol is quantified by the effective QFI defined in Eq. (\ref{eq:effQFIapp}). Being the quantum states Gaussian, also this figure of merit can be derived from the information contained in first and second moments. In particular, one shows that the homodyne classical Fisher information can be evaluated as \cite{Genoni2013PRA}
\begin{align}
     \mathcal{F}_{\sf hom} = \mathbbm{E}_{\sf traj}\left[2 \int dt \, (\partial_\omega \bar{\bf r}_c)^{\sf T} B B^{\sf T} (\partial_\omega \bar{\bf r}_c)\right] \,,
    \label{eq:FhomApp}
\end{align}
while the QFI of the (Gaussian) conditional state is obtained via the formula \cite{Pinel2013}
\begin{align}
    \mathcal{Q}[\varrho_c] &= \frac{\Tr\left[\left(\boldsymbol{\sigma}_c^{-1} (\partial_\omega \boldsymbol{\sigma}_c) \right)^2 \right]}{2(1+ \mu^2)} +\frac{2 (\partial_\omega \mu)}{1-\mu^4} \, \nonumber \\
    &\,\,\,\,
    + 2  (\partial_\omega \bar{\bf r}_c)^{\sf T} \boldsymbol{\sigma}_c^{-1} (\partial_\omega \bar{\bf r}_c) \,,
\end{align}
with 
\begin{align}
\mu = \Tr[\varrho_c^2] = 1/\sqrt{{\rm Det}[\boldsymbol{\sigma}_c]}
\label{eq:purity}
\end{align}
denoting the purity of the conditional quantum state. 
We thus also need the evolution of the derivatives of first and second moments respect to the parameter $\omega$, that can be numerically integrated via the equations \cite{Genoni2013PRA}
\begin{widetext}
\begin{align}
    d(\partial_\omega \bar{\bf r}_c) &= (\partial_\omega A) \bar{\bf r}_c \,dt + A (\partial_\omega \bar{\bf r}_c) \,dt - \frac{(\partial_\omega \boldsymbol{\sigma}_c) B \, d{\bf w}_t}{\sqrt{2}} + (E - \boldsymbol{\sigma}_c B) B^{\sf T} (\partial_\omega \bar{\bf r}_c) \, dt 
        \label{eq:drc}
    \\
    \frac{d (\partial_\omega \boldsymbol{\sigma}_c) }{dt} &= (\partial_\omega A) \boldsymbol{\sigma}_c + \boldsymbol{\sigma}_c (\partial_\omega A) ^{\sf T} + A (\partial_\omega \boldsymbol{\sigma}_c) +(\partial_\omega \boldsymbol{\sigma}_c) A^{\sf T} + (\partial_\omega \boldsymbol{\sigma}_c) B (E - \boldsymbol{\sigma}_c B)^{\sf T} + (E - \boldsymbol{\sigma}_c B) B^{\sf T} (\partial_\omega \boldsymbol{\sigma}_c) \,.
    \label{eq:dsigmac}
\end{align}
\end{widetext}
At each time $t$, also the purity $\mu$ and its derivative $\partial_\omega\mu$ can be directly obtained from $\boldsymbol{\sigma}_c$ and $(\partial_\omega \boldsymbol{\sigma}_c)$ via the formula in Eq. (\ref{eq:purity}).

\par
We remark that these quantities are also exploited as {\em observations} for the neural network that optimizes the feedback strategy. In order to train our agent and to assess the performance of the different protocols, we have thus numerically simulated different trajectories of the quantum states via Eqs. (\ref{eq:rcApp}), (\ref{eq:sigmacApp}), (\ref{eq:drc}) and (\ref{eq:dsigmac}), and we have performed the numerical integral and the numerical average in Eqs. (\ref{eq:FhomApp}) and (\ref{eq:effQFIapp}).
\section{The effect of the Hamiltonian coupling constant $\chi$}\label{a:coupling}
\begin{figure*}[t]
    \centering
    \includegraphics[width=.95\textwidth]{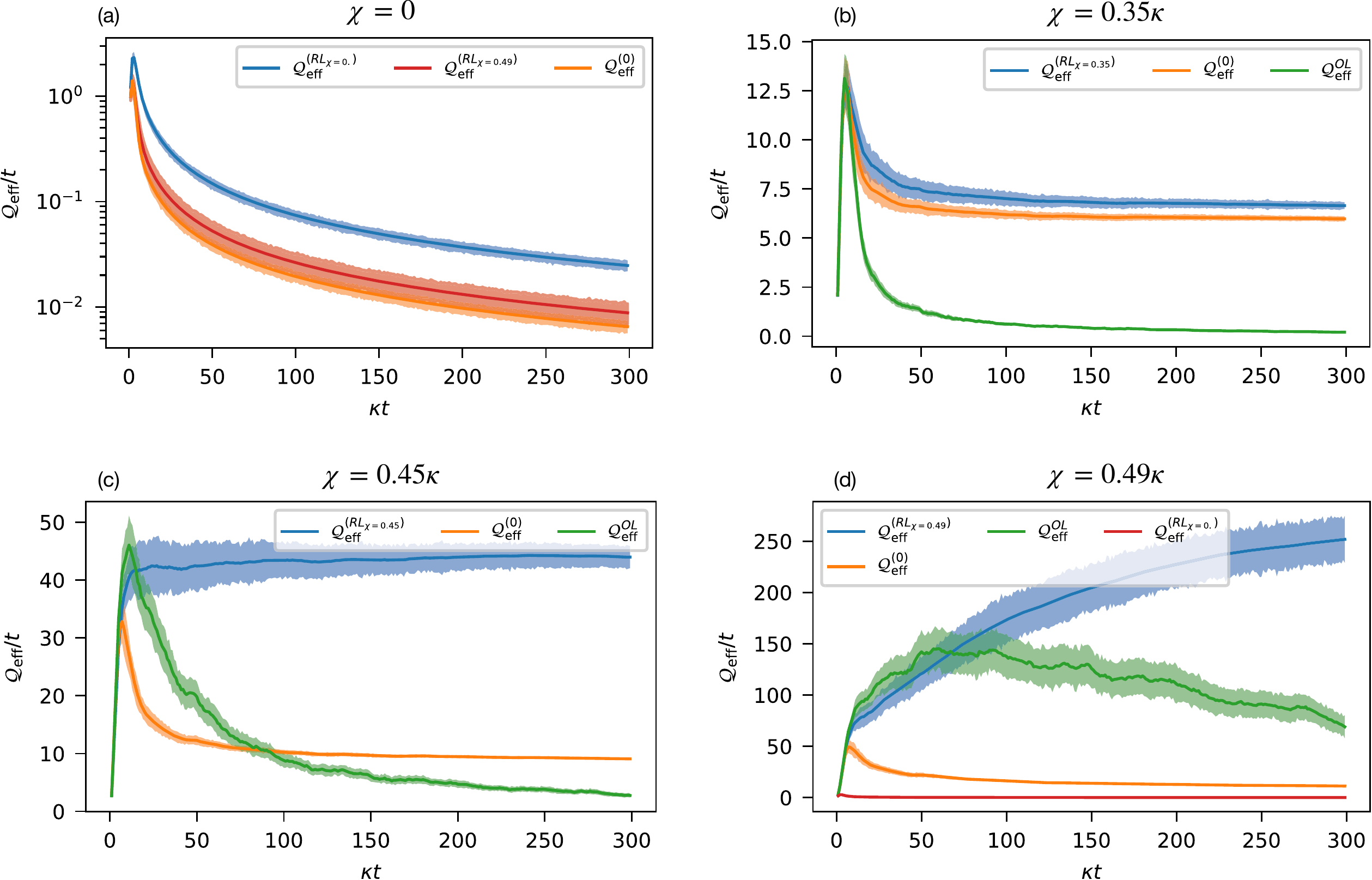}
    \caption{Performance of the control strategies as a function of time, quantified by the effective quantum Fisher information $\mathcal{Q}_{\sf eff}$ divided by time, and for different values of $\chi$: panel (a): $\chi=0$; panel (b) $\chi=0.35\kappa$; panel (c): $\chi=0.45\kappa$; panel (d): $\chi=0.49\kappa$. Notice that in panel (a) the scale of the y-axis is logarithmic and that the curve corresponding to the open-loop strategy $\mathcal{Q}_{\sf eff}^{(OL)}$ has not been reported as the corresponding values are almost negligible.  
    Furthermore in panel (a) and panel (d) we have added an extra red curve, corresponding to the agent trained at respectively $\chi=0.49$ and at $\chi=0$.\\
    The results have been obtained simulating $N=1000$ trajectories with a time-step $dt=0.001/\kappa$, by fixing the other parameters as $\omega=0.1\kappa$, $\eta=0.9$ and by considering as an initial state a thermal state with $n_{\sf th} = 5$ and initial first moment vector $\bar{\bf r}_c(0) = (0,0)$.}
    \label{fig:varyingchi}
\end{figure*}

In this appendix we discuss the role of the Hamiltonian coupling constant $\chi$ in the learning of the strategy by the agent, and on its performances. As we highlighted in the text, the parameter $\chi$ is directly responsible for the generation of squeezing in the conditional states. In fact squeezing can be observed if and only if $\chi$ is larger than zero, and in particular near criticality, i.e. for $\chi \approx \kappa/2$ the amount of squeezing generated is close to infinity in the case of perfect monitoring. \\
In Fig. \ref{fig:varyingchi} we plot our figure of merit $\mathcal{Q}_{\sf eff}/t$, obtained by agents trained with different values of $\chi=\{0 ,0.35\kappa,0.45\kappa, 0.49\kappa \}$, as a function of time, and compare it with the other benchmark strategies. We observe how the agent allows to reach values of QFI larger than those obtained by means of the other strategies, in particular in the long time limit. Other relevant observations can be drawn from these plots: 
i) In general, we find that, for all strategies, larger values of $\chi$ yield larger values of QFI thus highlighting, once again, the importance of the squeezing generated during the dynamics. ii) For values of $\chi$ that are {\em large enough} (e.g. for $\chi=0.45\kappa$ and $\chi=0.49\kappa$) the agent is able to devise a strategy such that the maximum of $\mathcal{Q}_{\sf eff}/t$ is observed in the long time limit; for smaller values of $\chi$ (e.g. $\chi=0.35\kappa$) one observes a maximum at short times, while in the long time limit the ratio between effective QFI and time tends to a smaller stationary value. iii) We also find that for $\chi=0.45\kappa$ the maximum of $\mathcal{Q}_{\sf eff}/t$ obtained for the open-loop control strategy is compatible with the value obtained in the long time limit via the agent's strategy. We stress, however, that while for the open-loop strategy one would need to stop the dynamics at a very specific time, by employing the RL-strategy one has a wide available time window, in which the dynamics has reached a steady-state behaviour.

To better describe the properties of the strategies devised by the neural network, we have added an extra curve the in the panels (a) and (d) of Fig. \ref{fig:varyingchi}: in panel (a) we have added the values of QFI corresponding to the agent trained at $\chi=0.49\kappa$ applied to the case $\chi=0$, while in panel (d) we have added the performance of the agent trained at $\chi=0$ applied to the case $\chi=0.49\kappa$. In the first scenario we observe that the agent trained at $\chi=0.49\kappa$ is still able to beat the benchmark strategies and is only slightly less performing than the properly trained agent; in the second scenario the situation is completely inverted: the agent trained at $\chi=0$ is indeed yielding very low values of QFI. Our interpretation of this result is the following: when the agent is trained near criticality, it learns how to optimize both first moments and squeezing, and thus is performing well also when squeezing is absent. On the other hand, at $\chi=0.49\kappa$, that is when squeezing is playing a major role in the estimation protocol, the agent trained in the no-squeezing scenario completely fails in enhancing the estimation precision.
\section{The effect of continuous monitoring efficiency}\label{a:efficiency}
Here we present some results that we obtained for different values of the monitoring efficiency, $\eta = \{0.9, 0.5, 0.1\}$. 
In Fig. \ref{f:compareQeffEta} we report the behaviour of the different Fisher information divided by time as in Fig.  \ref{f:compareQeff}. We observe how the enhancement is still clearly observed for $\eta=0.5$ and that even for very small monitoring efficiency ($\eta=0.1$) the agent-feedback is able to yield a larger estimation precision compared to the other strategies, in particular in the long time limit. \\

We have also reported in Fig. \ref{f:scatterSqFM} scatter plots of the perpendicular squeezing defined in the main text and the absolute value of the first moment vector $|\bar{\bf r}_c|$ for a fixed time $\kappa t = 180$. We observe as remarked also in Fig. \ref{f:perpsqueezing}, that, with respect to the {\em no-control} strategy, for $\eta=0.9$ the agent is able to prepare trajectories with larger perpendicular squeezing. However in this plot we also observe that the agent is in general able to prepare conditional states that, for a fixed amount of perpendicular squeezing, yields larger first moments, and thus leading to an enhanced estimation. By reducing the monitoring efficiency, we find that the first effect (trajectories with larger perpendicular squeezing) is basically lost also for $\eta=0.5$, while the second effect, that is, larger first moments at fixed squeezing, is still obtained and thus it can be considered as the sole responsible for the better estimation precision. 

The relevance of the first moment vector is also highlighted in Fig. \ref{f:scatterFMfHom}, corresponding to the scatter plot of $|\bar{\bf r}_c|$ and $\log (f_{\sf hom}^{\sf (traj)})$ for the different trajectories, where we have introduced the quantity 
\begin{align}
    f_{\sf hom}^{\sf (traj)} = 2 \int dt \, (\partial_\omega \bar{\bf r}_c)^{\sf T} B B^{\sf T} (\partial_\omega \bar{\bf r}_c) \,, \label{eq:fhom}
\end{align}
corresponding to the contribution of each trajectory to the homodyne Fisher information $\mathcal{F}_{\sf hom} = \mathbbm{E}[ f_{\sf hom}^{\sf (traj)}]$ (see Eq. (\ref{eq:Fhom})). In the figure we indeed observe how the two quantities seem to be correlated for the agent strategy, while they seem uncorrelated for the {\em no-control} strategy (we remind that for the OL-strategy $f_{\sf hom}^{\sf (traj)}=0$ for each trajectory), highlighting once again the mechanism behind the strategy devised by the agent.

\section{On the effectiveness of homodyne detection as a final strong measurement}\label{a:finalhomodyne}
In this appendix we discuss the effectiveness of homodyne detection as a final strong measurement in our protocol, i.e., we calculate the FI of a final homodyne measurement and we compare it to the QFI  $\bar{\mathcal{Q}}_c$. We remark that in general a non-Gaussian measurement may be needed to saturate the quantum Cram\'er-Rao bound, that is to obtain a classical FI equal to the QFI; however we expect that homodyne detection is going to extract a fair amount of the maximum information achievable, being nearly optimal for pure Gaussian states (see Refs.~\cite{Monras2006,Oh2019} for an extensive study of phase estimation with Gaussian states).

A projective Gaussian measurement can be modeled by the covariance matrix of a squeezed vacuum state, with squeezing parameter $s$, and a phase rotation of angle $\theta$:
\begin{equation}
    \boldsymbol\sigma_m(z, \theta) = R(\theta) \text{diag}(z, 1/z) R(\theta)^{\sf T},
\end{equation}
where $z = \exp(2s)$ and $R(\theta)$ is a rotation matrix. In the limit $z \rightarrow 0$ we have a homodyne measurement: when $\theta = 0$ ($\theta = \pi /2)$, the quadrature $\hat q$ ($\hat p$) is measured.

The measurement outcome probability of such measurement on a state with first moments $\mathbf{r}_c$ and covariance matrix $\boldsymbol{\sigma}_c$ is a bidimensional Gaussian distribution $\mathcal{N}(\mathbf{r}_c, \Sigma)$, where $\Sigma = (\boldsymbol\sigma_c + \boldsymbol\sigma_m)/2$, for which the FI reads
\begin{equation}
    \mathcal{F}[\mathcal{N}(\mathbf{r}_c, \Sigma)|\omega] = (\partial_\omega \boldsymbol{r}_c)^{\mathsf{T}} \Sigma^{-1} \partial_\omega \boldsymbol{r}_c
  + \frac 12 \text{Tr}[(\Sigma^{-1} \partial_\omega \Sigma)^2].
\end{equation}

\begin{figure}[t]
    \centering
    \includegraphics{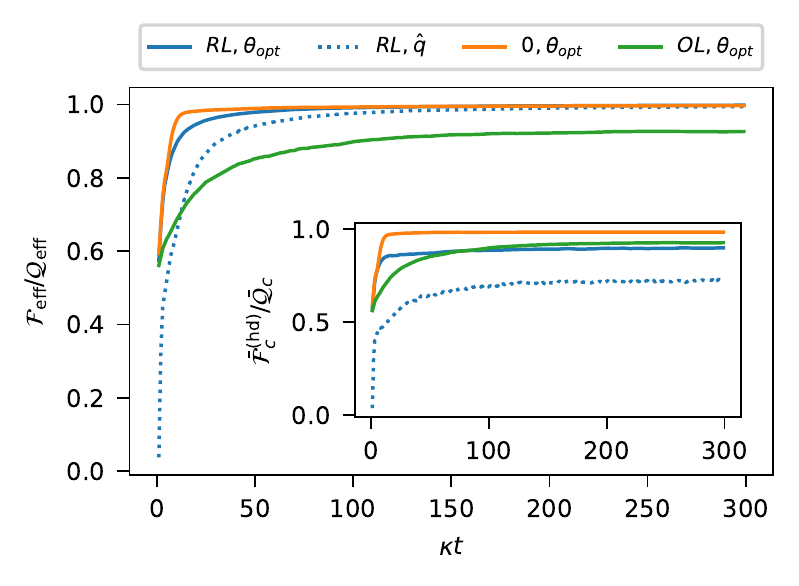}
    \caption{Ratio between the effective FI  $\mathcal{F}_{\sf eff}$ for final homodyne detection and the effective QFI of the conditional state $\bar{\mathcal{Q}}_{\sf eff}$ as a function of $\kappa t$ for the three different strategies (the inset shows the ratio between the average FI of the strong homodyne measurement $\bar{\mathcal{F}}_c^{(\sf hd)}$ and the average QFI $\bar{\mathcal{Q}}_c$. The solid lines show the FI $\bar{\mathcal{F}}_{\sf hd, opt}$ optimized over $\theta$, while the dotted line shows the FI $\bar{\mathcal{F}}_{\sf hd, \hat q}$ obtained with a homodyne on the quadrature $\hat q$ for the RL strategy. In all cases, while not being optimal, homodyne detection allows to extract a significant fraction of the total available information on the frequency $\omega$ (parameters values are fixed as in Fig.~\ref{f:compareQeff}).}
    \label{fig:fi_qfi_ratio}
\end{figure}

For each trajectory, we maximize the FI for a final homodyne measurement over $\theta \in [-\pi/2, \pi/2]$, $\mathcal{F}_c^{(\sf hd)}$. We then calculate the average $\bar{\mathcal{F}}_c^{(\sf hd)}$ and compare it with
$\bar{\mathcal{Q}}_c$. The resulting ratio $\bar{\mathcal{F}}_c^{(\sf hd)}/\bar{\mathcal{Q}}_c$ is shown in Fig.~\ref{fig:fi_qfi_ratio} for the three different strategies. As can be seen, the optimized homodyne detection, although not ideal, allows for the extraction of a significant amount of information from the quantum state. Finding the optimal angle $\theta$ in real-time should be possible by means, for example, of a FPGA (see discussion on the feedback real-time implementation in Sec.~\ref{s:conclusion}). The dotted lines in the inset of Fig.~\ref{fig:fi_qfi_ratio} show that even if a fixed, and suitably a priori chosen, angle $\theta$ is used,  it is anyway possible to extract a significant fraction of the QFI for the state (we set $\theta = 0$ for the RL strategy, corresponding to measuring $\hat q$, as suggested by a direct inspection of the distribution of the optimal $\theta$ values for the different trajectories). The overall effect on the effective Fisher information when using homodyne detection as strong final measurement is shown in the main panel of Fig.~\ref{fig:fi_qfi_ratio}: in the strategy devised by the RL agent, where the contribution of the final strong measurement to $\mathcal{Q}_{\sf eff}$ is small, the ratio between the effective Fisher information $\mathcal{F}_{\sf eff}$ corresponding to a final homodyne detection (either optimized or for $\theta=0$) and the corresponding to effective QFI $\mathcal{Q}_{\sf eff}$ tends rapidly to one, by increasing the monitoring time $\kappa t$. We also notice that the ratio is reasonably high also for the other benchmark strategies: the more significant effect can be seen in the OL scenario, particularly at short times, due to the fact that $\mathcal{Q}_{\sf eff} = \bar{\mathcal{Q}}_c$.

\begin{figure}[t]
    \centering
    \includegraphics[width=\columnwidth]{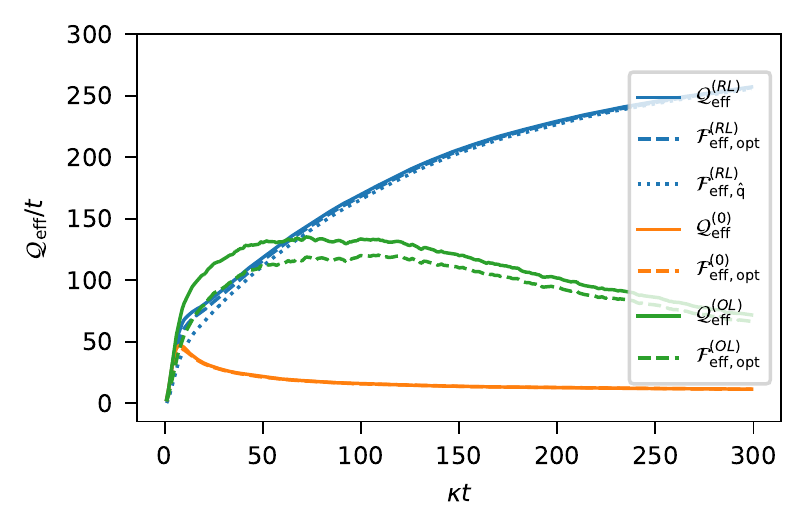}
    \caption{$\mathcal{Q}_{\sf eff}/t$ (solid lines) and $\mathcal{F}_{\sf eff}/t$ for homodyne detection (dashed lines) as functions of $\kappa t$ for the three different strategies under consideration. Using homodyne detection as a final strong measurement does not affect the precision in the estimation significantly for the RL strategy, given that the biggest contribution to $\mathcal{Q}_{\sf eff}$ comes from the continuous monitoring. The dotted blue line shows $\mathcal{F}_{\sf eff}/t$ for homodyne detection on the $\hat q$ quadrature, for the RL strategy. Even without optimizing the angle of the final homodyne measurement, the RL-strategy allows to overcome the performances of the other benchmark strategies
    (parameters values are fixed as in Fig.~\ref{f:compareQeff}).}
    \label{fig:qeff_hd_compare}
\end{figure}

We finally compare the performance of a final homodyne detection for the three strategies in Fig.~\ref{fig:qeff_hd_compare}, where we indeed observe that the enhancement obtained via the RL-strategy is still mantained when a final homodyne detection is performed.  


\end{document}